\documentclass[]{aa}
\usepackage{graphicx}
\usepackage[authoryear]{natbib}
\bibpunct[,]{(}{)}{;}{a}{,}{,}

\newcommand{\msolyr}{\ifmmode{{\rm M}_{\odot}~{\rm yr}^{-1}}
\else{{M$_{\odot}$~yr}$^{-1}$}\fi} 
\newcommand{\msun}{\ifmmode{{\rm
M}_\odot} \else{M$_{\odot}$} \fi} 
\newcommand{\lsun}{\ifmmode{{\rm
L}_{\odot}}
\else{L$_{\odot}$} \fi}
\newcommand{\rsun}{\ifmmode{{\rm R}_{\odot}}
\else{R$_{\odot}$} \fi}
\newcommand{\zsun}{\ifmmode{{\rm Z}_{\odot}}
\else{Z$_{\odot}$} \fi}
\newcommand{\teff}{\ifmmode{{\rm T}_{\rm eff}}
\else{T$_{\rm eff}$} \fi}
\newcommand{\mdot}{\ifmmode{\dot{\rm M}}
\else{$\dot{\rm M}$}\fi}

\begin{document}

\title{Low and intermediate mass star yields:
 The evolution of carbon abundances}

\author{Marta Gavil\'{a}n\inst{1}, James F. Buell\inst{2}\footnote{New
 adress:Department of Math and Physics, SUNY Alfred, Alfred, NY 14802,
 USA, e--mail:buelljf@alfredstate.edu} and Mercedes Moll\'{a} \inst{3}}

\offprints{Marta Gavil\'{a}n}

\institute{Departamento de F\'{\i}sica Te\'{o}rica,
Universidad Aut\'onoma de Madrid, 28049 Cantoblanco, Spain\\
\email{mgavilan@eresmas.net}\\
\and Department of Physics, Northeastern State University,
Tahlequah, OK 74464, USA \\ \email{buell@cherokee.nsuok.edu} \\
\and Dpto. de F\'{\i}sica de Fusi\'{o}n y Part\'{\i}culas elementales, 
C.I.E.M.A.T., Avda. Complutense 22, 28040 Madrid, Spain 
\\ \email{mercedes.molla@ciemat.es}}

\date{Received ; accepted }

\titlerunning{The evolution of carbon abundances}
\authorrunning{Gavil\'{a}n, Buell \& Moll\'{a}}

\abstract{ We present a set of low and intermediate mass star yields
based on a modeling of the TP--AGB phase which affects the production
of nitrogen and carbon. These yields are evaluated by using them in a
Galaxy Chemical Evolution model, with which we analyze the evolution
of carbon abundances. By comparing the results with those obtained
with other yield sets, and with a large amount of observational data,
we conclude that the model using these yields combined with those from
\citet{woo95} for massive stars properly reproduce all the data. The
model reproduces well the increase of C/O with increasing O/H
abundances. Since these massive star yields do not include winds, it
implies that these stellar winds might have a smoother dependence on
metallicity than usually assumed and that a significant quantity of
carbon proceeds from LIM stars.

}

\maketitle

\section{Introduction}

The study of galactic evolution gives important clues about the
Universe. The chemical evolution of a galaxy depends mainly on three
factors: a) the Initial Mass Function (IMF), b) the Star Formation
Rate (SFR), and c) the products ejected to the interstellar medium as
a consequence of the stellar evolution. Due to this last ingredient, it is
quite important to know the mechanisms of element production in the
interior of the stars, where they are produced, in what quantity, and
when they are ejected to the interstellar medium. The study of these
processes is made through the stellar yields, a concept introduced by
\citet{tin80}. Since then, a lot of work has been done in this
field to obtain and improve them, first for solar metallicity and
then for other stellar metallicities.

Modern studies \citep[][hereinafter CHIA03] {car00,lia01,chia03} have
shown that differences in stellar yields reflect appreciable
variations in the results obtained with Galaxy Chemical Evolution
models.  One of the reasons for those differences is our still poor
understanding of the evolution of low and intermediate mass (LIM)
stars in the post main sequence stages. Another reason is the poor
knowledge of the influence of the metallicity on the mass loss by
stellar winds.  Thus, in spite of the large effort to compute yields,
the matter is not clear at all: any improvement in the stellar
evolution theory may have an effect on yields, and, therefore, on
chemical evolution model results.

For massive stars, one of the most frequently used set of stellar yields has been
the one calculated by \citet[][ hereafter WW]{woo95}. They added to
the pre--supernova yields \citep{woo86} the nucleosynthesis elements produced in supernova
type II explosions, for metallicities between $Z = 0$ and
Z$_{\odot}$. \cite{mae92} calculated the yields for stars with mass
between 1 M$_{\odot}$ and 120 M$_{\odot}$, for metallicities in the
range from $Z = 0.0001$ to Z$_{\odot}$, as proceeding from the stellar
winds which occur during stellar evolution.  Then, 
\citet[][hereafter PCB]{pcb98} obtained massive star yields for the
metallicities of the Padova group taking into account all these
ingredients: the loss of mass by stellar winds, the influence of the
metallicity during stellar evolution and the explosive
nucleosynthesis, for computing a complete set of yields.  More
recently, \cite{limongi} and \cite{raus02} have calculated new massive
stars yields, but the mass and/or metallicity range is not so wide as
the {\sl old } yield sets from WW and PCB. As we would like to
compare our results with other works we choose only those {\sl
old } yield sets, widely used in the literature.

  For the low and intermediate mass star yields, one of the most used
  sets has been the one obtained by \cite{rv81} (
  hereafter RV), calculated for LIM stars taking into account the
  effects of convective dredge--up and inner layer burning,
  the so--called {\sl Hot Bottom Burning} processes.  However, these
  yields are not successful in predicting the observed abundances C/O
  and N/O. Some other works such as \citet{mar96,for97,mar98} and \citet[][
  hereafter VG]{vhoek97}, or, more recently, \citet[][ hereafter
  MA]{mar01} have treated the evolution of these last phases in this
  kind of star. These works compute new sets of stellar yields,
  which, however, show a very different behavior.

This is the reason why new yields for LIM stars were recalculated 
\citep{buell97} using the same basic scheme as RV.  Calculations 
shown and used in this work are based on the study of the
transformations that stars with masses between 0.8 and 8 \msun suffer
after the Main Sequence, mainly on the thermally pulsing asymptotic
giant branch (TP--AGB), when the third dredge---up (TDU) and Hot Bottom
Burning (HBB) processes take place.  As these processes affect mainly
the production of carbon and nitrogen, these yields will have a
different behavior in these elements to that found by other authors. We
present this new set of yields for metallicities between
$log(Z/Z_{\odot})=-0.2 $ (or $Z = 0.0126$) and $log(Z/Z_{\odot})=+0.2$
($Z = 0.0317$). They are calculated taking into account, as is commonly
accepted, that the LIM stars produce primary and secondary components
for the CNO elements, and, therefore, these two components are given
separately.  Besides that, we evaluate these yields by using them as
input in a Galactic Chemical Evolution model, and comparing the
results with those obtained with other published yield sets, in
particular with RV and VK.

There are some recent works \citep[] {ven02,dray03}
describing detailed evolutionary computations followed from the pre--MS
phase up to the very late evolutionary stages for LIM stars, including
in particular the AGB phase, and giving the element yields. The
problem with using these new sets resides in the incomplete range in
mass and Z, necessary for our purposes, due to the large computation
time necessary to calculate those models compared with the
corresponding one using less exact methods. As we
want to compare the results with those obtained by other authors, 
we prefer to use VG and MA. We check, whenever
possible, if differences between our stellar yields and those obtained
with more precise methods are significant.

From the observational point of view, there are several open questions
about the primary or secondary character of nitrogen and carbon, which
have remained unsolved up to now. In particular, for the carbon
abundances, the graph of log(C/O) {\sl vs} O/H shows first a flat line
which then increases for oxygen abundances larger than $12+log(O/H)
=8$. The flat slope is usually explained as being due to the C mostly being
ejected by massive stars with oxygen. However, it is difficult to explain
the posterior increase of C/O on the basis of a primary behavior for the two
elements.  \cite{car00,hen00} and other authors claim that the
dependence on metallicity of yields, due mostly to its effect on mass
loss by stellar winds in massive stars, is essential for solving this
problem, while some others try to explain how a secondary element,
proceeding from the carbon ejected by LIM stars, can show this kind of
behavior. Here we use new LIM star yields and we will probe their
effect on the carbon abundance.

We present the new low and intermediate mass yields in Section
2. Section 3 is devoted to the comparison of the different yield sets
used for our purposes, and to explain how to calculate the data sets
used as inputs to the chemical evolution model. In Section 4 we
summarize the multiphase chemical evolution model and show the
resulting calibration for the Milky Way galaxy (MWG). Then the
corresponding results for carbon abundances are given.  The discussion
is given in Section 5 and our conclusions are presented in Section 6.

\section{The stellar evolution of LIM along the TP--AGB phase}  

The evolution after the main sequence (MS) of stars with masses
between 0.8 and 8 \msun is determined primarily by the value of their
ZAMS mass. The different stages of evolution are described according to
the following mass ranges:

\begin{itemize}

\item{0.8 M$_{\odot}$ -- 1.7 M$_{\odot}$: During thermal pulses the
convective envelope does not penetrate deeply enough to mix processed
material into the surface layers, and the composition before and after
thermal pulses are the same. The temperature at the base of the
envelope between thermal pulses does not get high enough for HBB to
occur, therefore there is no change in the composition of these stars
during the TP--AGB. They experience a 1$^{\rm st}$ dredge--up event
which modifies the composition of the ejected material. They do not
suffer 2$^{\rm nd}$ or 3$^{\rm rd}$ dredge--up events.} 

\item{1.7 M$_{\odot}$ -- 4  M$_{\odot}$: Abundances in these stars are
dominated by  the 3$^{rd}$ dredge--up, because they  undergo many events
of this  type. The temperature at  the base of their  envelopes is not
high enough to allow HBB, and therefore each dredge--up increases the C
and He abundances. The abundance of nitrogen is increased by the first
dredge--up.}

\item{4 M$_{\odot}$ -- 8 M$_{\odot}$: The envelope bases for 4
M$_{\odot}$ stars reach a temperature of $\ge 30\times 10 ^{6}$ K, and
then HBB -- a CNO cycle at the base of the convective envelope --
can take place. The consequence is a decrease of the
abundance of carbon and an increase of nitrogen abundance. If there
is enough time, it is even possible to convert $^{16}O$ to $^{14}N$,
thus producing a decrease in the oxygen abundance. The ratio N/O will
show a severe increase for stars around 4 M$_{\odot}$, when the HBB
process begins. These stars also typically experience a 2$^{\rm nd}$
dredge--up which increases their surface helium abundance.}
\end{itemize}

Only a brief summary of the inputs into LIM star models used will be
presented here, mostly those details that significantly affect the
calculated yields. Additional details can be found in \cite{buell97}
and \cite{buell04}

\subsection{Luminosity}

The luminosity of a TP--AGB star immediately after the first thermal
pulse starts at a low value and then grows rapidly during successive
pulses until it reaches an asymptotic value. This asymptotic value can
be expressed as a function of the core mass and the mass of the star. 
The asymptotic luminosity of the low--mass stars on the thermally pulsing
asymptotic giant branch (TP--AGB) is usually modeled using the
core--mass--luminosity relations of \cite{bs92}.

In recent years, however, it has become apparent that a simple
core--mass luminosity relationship (with or without a metallicity
dependence) is not appropriate for intermediate mass stars
(M$>3.5\msun$). The luminosity now appears to depend on the stellar
mass as well. \cite{tgb83} showed using semi--analytic arguments that a
core--mass luminosity relation holds for AGB stars only when the
hydrogen burning shell is separated from the convective envelope. They
found that a core--mass luminosity relationship is not appropriate if
the convective shell penetrates the hydrogen burning
layer. \cite{bs91} modeled a 7$\msun$ star and found that it did not
follow any kind of core--mass luminosity behavior because the
convective envelope penetrated the hydrogen burning layer. This effect
has been confirmed by the TP--AGB models of
\cite{bs92,l92,bsa93,vas93,b95,for97}, and \cite{str00}.

The asymptotic value of the surface luminosity for stars of all masses
is found from:

\begin{equation}
L_s =f{\rm L}_{\rm cm} 
\end{equation}

where

\begin{eqnarray}
f&=&1+0.186({\rm M}-2.17) , \,\, {\rm M} > 2.17\msun \\
f&=&1                     , \,\, {\rm M}\le 2.17\msun 
\end{eqnarray}

M is the total mass of the star ${\rm L}_{\rm cm}$ is the luminosity
if a core--mass luminosity relationship were followed, and $f$ is a
factor to correct the luminosity for the effects of HBB. This
relationship was derived by fitting a function to the AGB models of
\cite{bs92} and \cite{bsa93}.  The luminosity depends strongly on mass
over 4$\msun$.

For low--mass stars (${\rm M}<3{\rm M}_{\odot}$) we adopted the
relationship from \cite{bs88}:

\begin{eqnarray}
{\rm L}_{\rm cm}&=&238000\mu^3({\rm Z}_{\rm CNO})^{0.04}({\rm M}_{\rm c}^2-0.0305{\rm M}_{\rm c}-0.1802).
\end{eqnarray}

where ${\rm Z}_{\rm CNO}$ is the mass fraction of carbon, nitrogen,
and oxygen and $\mu$ is the mean molecular weight of the envelope. 
This relationship approximates the metallicity variations of the luminosity.

For higher mass stars (${\rm M}\ge3{\rm M}_{\odot}$) we adopted the
following relationship for the core--mass luminosity relationship:

\begin{eqnarray}
{\rm L}_{\rm cm}&=&52000({\rm M}_{\rm c}-0.456).\\
\end{eqnarray}

Core--mass luminosity relations only give the luminosity at the local
``asymptotic'' limit. It is well known that the luminosity during the
first inter--pulse does not correspond to the core--mass luminosity
relation, and in general 5--10 pulses are needed to reach it. The first
thermal pulses occur when the helium burning shell still produces a
significant fraction of the luminosity ($\sim$50\%), but after a
few pulses the helium burning shell only produces a few percent of the
luminosity.  

As mentioned earlier, the luminosity at the first pulse is less than
the asymptotic value.  The value depends on the mass of the core at the
onset of the first pulse. There also appears to be an effect due to
the metallicity. For M$_{\rm c}>0.7\msun$ we adopt the following
relation for the luminosity at the first pulse;

\begin{equation}
\log{\rm L_s(0)}=2.07{\rm M}_{\rm c}+2.48-3(.02-Z)
\end{equation}

where Z is the metallicity of the model. This relation is a fit to the
models of \cite{bs92} and \cite{bsa93}. For models with M$_{\rm
c}\le0.7\msun$ the expressions of \cite{l86} are used:

\begin{eqnarray}
L(0)=29000(M_c-0.5)+1000&Z=0.001\\
L(0)=27200(M_c-0.5)+1300&Z=0.02,
\end{eqnarray}

where values at other metallicities are found by linearly 
extrapolating/interpolating in $\log{\rm Z}$.

The mass dependence of the luminosity is a consequence of hot--bottom
burning and determines whether a star yields mainly carbon or nitrogen.
Stars of lower mass (M$\le4\msun$) exhibit little or no hot bottom
burning and nearly all of the $^{12}{\rm C}$ mixed into the envelope
is ejected into the interstellar medium as $^{12}{\rm C}$. Stars of
higher mass (M$\ge4\msun$) exhibit efficient HBB effectively
converting all the $^{12}{\rm C}$ in the envelope into $^{13}{\rm C}$
and $^{14}{\rm N}$ until the envelope mass is reduced to a point where
HBB no longer occurs.

\subsection{Mass Loss}

On the AGB mass--loss rates are large enough to effect the evolution of
the star. The mass--loss rates are calculated from the following
formulas:

\begin{enumerate}

\item{The \cite{rei75} mass--loss rate:\\ $\dot{\rm M}_{\rm
R}=-(4\times 10^{-13})\eta{\rm L}{\rm R}/{\rm M}$.}

\item{The pulsation period mass loss rate of \cite{vas93}, $\dot{\rm
M}_{\rm PP}$, given by:\\

$\log{\dot{\rm M}}(\msolyr) = -11.4+0.0123{\rm P}$\\ 
with:\\

$\log{P~}(days) =  -2.07+1.94\log{R/R_{\odot}}-0.9\log{M/\msun} $

where R is the radius of the star. (Note, their modification for
M$>2.5\msun$ is not included)}.

\item{A superwind mass--loss rate, $\mdot_{\rm SW}$, which we take as
$5\times 10^{-5}\msolyr$.}
\end{enumerate}

The first relation is followed until $\mdot_{\rm PP}>\mdot_{\rm R}$, after
which relation 2 is used. Relation 2 is used until $\mdot_{\rm PP}>
%is it < instead > ? 
\mdot_{\rm SW}$, after which a constant mass--loss rate of $5\times
10^{-5}\msolyr$ is used.

This mass--loss prescription is metallicity dependent because the
mass--loss rate depends on the radius of the star. Because the radius
decreases as metallicity decreases so does the mass--loss rate. This
affects the yields by increasing the time a star spends on the TP--AGB
allowing for more dredge--up events to occur, increasing the amount of
primary production of either carbon or nitrogen in LIM. There is a
slight countering effect due to luminosity. Lower metallicity stars
usually end up on the TP--AGB with higher core masses than higher
metallicity stars and this increases the mass--loss rate. However, the
opacity effect appears to be the dominant effect on the mass--loss
rate.

\subsection{Third Dredge Up}

Between thermal pulses the base of the convective envelope and the
core of the star move outward in mass by an amount $\Delta{\rm M}_{\rm
c}$ and at the end of the thermal pulse which follows the convective
envelope can penetrate into this region and mix this modified material
into the envelope of the star. The depth to which the convective zone
penetrates is represented by the parameter $\lambda$. The mass of
material mixed into the envelope, ${\rm M}_{\rm dredge}$, is:

\begin{math} 
{\rm M}_{\rm dredge}=\lambda\Delta{\rm M}_{\rm    c}. 
\end{math}  
There have been several recent papers on TDU and the value of lambda
but no quantitative agreement. The mass dredged up depends on the
assumptions used. Most authors use convective overshooting
\citep{mow99,klp03,her97,her00}, which seems to indicate rather large
amounts of dredge--up. However, dredge--up can be obtained without it
\citep{str97}.

\begin{figure}
\resizebox{\hsize}{!}{\includegraphics[angle=0]{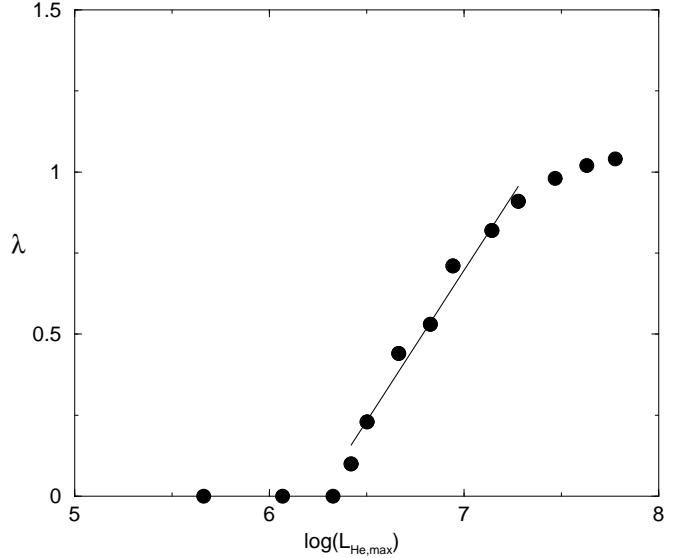}}
\caption{Plot of the value of the dredge--up parameter $\lambda$ versus
the maximum luminosity of the helium shell during a thermal pulse. The
data points are taken from Table 1 of \cite{her00}. The straight line
is a fit to value of lambda in pulses 4--10}
\label{lambda_fit}
\end{figure}

The parameter $\lambda$ is calculated using a formula from \cite{ba91}.
He showed that for dredge--up to occur the peak luminosity of the
helium burning shell during the shell flash, ${\rm L}_{\rm He,max}$,
must exceed a certain minimum, ${\rm L}_{\rm He,min}$, which is
dependent on stellar mass. Thus, we use his formula 
for the dredge up parameter:

\begin{equation}
\lambda=0.90(\log{{\rm L}_{\rm He,max}}-\log{{\rm L}_{\rm He,min}})
\end{equation}

with the constraint $0\le\lambda <1$. The formulas for both 
${\rm L}_{\rm He,max}$ and ${\rm L}_{\rm He,min}$ can be found in
\cite{buell97}. Bazan derived this from the TP--AGB models without
convective overshooting.

Most recent models have used convective overshooting to get TDU, however,
a qualitatively similar scheme can be derived for these models.  In
Figure~\ref{lambda_fit} we have plotted $\lambda$ versus the helium
luminosity for the 3\msun model of of \cite{her00}. We then fit the
value of lambda for pulses 4--10 with a linear curve. The equation of
the fit is:

\begin{equation}
\lambda=0.92887634(\log{{\rm L}_{\rm He,max}}-\log{{\rm L}_{\rm He,min}})
\end{equation}

The only significant difference between these equations is the
value of $\log{{\rm L}_{\rm He,min}}$ which is not surprising
considering one was derived from models without convective overshooting
and the other with. The values of $\log{{\rm L}_{\rm He,min}}$
for a 3\msun model with and without overshooting are respectively 6.25
and 7.08. This shows that it is more difficult to get dredge--up 
without overshooting.

\subsection{Abundance calculations}

The material mixed into the envelope is composed of about 75\%
helium, 24\% carbon-12, and 1\% oxygen-16 by mass. The abundance of
this material is calculated from the formulas in \cite{rv81}.

The program to calculate the LIM yields follows the surface abundances
of H, He, C, N, and O from the first thermal pulse to planetary nebula
ejection. The structure of the envelope during the time between
thermal pulses is calculated by solving the equations of stellar
structure for the convective envelope. The luminosity is determined as
a function of the stellar mass, the mass of the hydrogen--exhausted
core and the chemical composition. The effective temperature of the
stellar envelope is calculated by iterating this temperature until the
base of the convective envelope reaches the top of the hydrogen
exhausted core. The changes in abundances of H, He, C, N, and O in the
envelope due to nuclear reactions are computed using this structure.

The 1995 updated OPAL opacities ($\kappa _{OPAL}$) which are described
in \cite{ri92} were used when T$>$10000~K and the molecular opacities
($\kappa _{\rm mol}$) of \cite{af94} were used for T$<$6000~K. At
intermediate temperatures the opacity was computed by a weighted
average of both opacity sets. The abundance at the first thermal pulse,
determined by the effects of the first and second dredge--ups, is
computed using the formulas found in \cite{gdj93} and \cite{bi80},
respectively.

\subsection{Tuning the models}

The models were tuned by varying the mixing length parameter $\alpha$
until they fit a set of galactic planetary nebula abundances. The
value of the mixing length controls the radius of the star which in
turn controls the mass loss rate.

\begin{figure}
\resizebox{\hsize}{!}{\includegraphics[angle=-90]{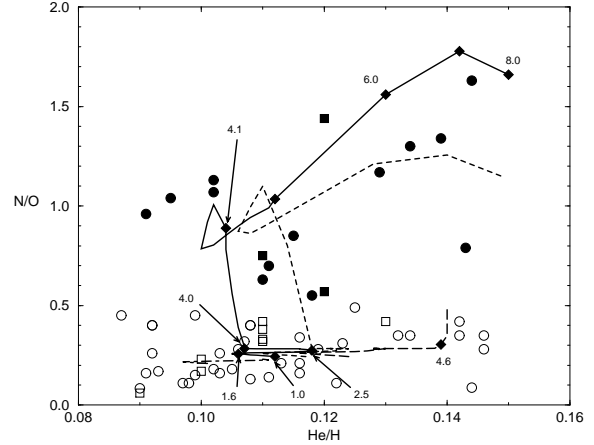}}
\caption{The circles and squares represent the KB and KH data sets,
respectively. Open and closed symbols respectively indicate PNe
with N/O$\le$0.5 and N/O$>$0.5, respectively. The solid, dashed,
long--dashed, and dash--dotted lines refer to models calculated
respectively with [Fe/H]=0.0, 0.1, 0.2, and -0.5. The mixing length
parameter, $\alpha$, of each model was set to 2.3. Only models with
M$<$4.6M$_{\odot}$ are shown for the [Fe/H]=0.2 and only models with
M$<$2M$_{\odot}$ are shown for the [Fe/H]= -0.5. The solid diamonds
indicate the results of models with [Fe/H]=0.0 and masses of 1.0, 1.6,
2.5, 4.0, 4.1, 6.0, and 8.0 M$_{\odot}$. Also the position of the
model with [Fe/H]=0.2 and mass of 4.6 M$_{\odot}$ is indicated.} 
\label{fig:heno_comp_all} 
\end{figure}
 
For a comparison between the PNe data and our models, we have chosen
two data sets because both have carbon abundances determined from IUE
data: 
\begin{enumerate} 
\item{The set described in
\citet[][hereinafter KH]{kh96} and references therein. This set
contains objects for which the abundances of helium, nitrogen, oxygen,
neon, and especially carbon have been carefully determined.} 
\item{The
sample of \citet[][hereinafter KB]{kb94} which contains 80 southern
Galactic PNe, for which the abundances of helium, nitrogen, oxygen,
neon, sulfur, and argon were determined. For some PNe the abundance of
carbon has also been determined.} 
\end{enumerate} 

\begin{figure} 
\resizebox{\hsize}{!}{\includegraphics[angle=0]{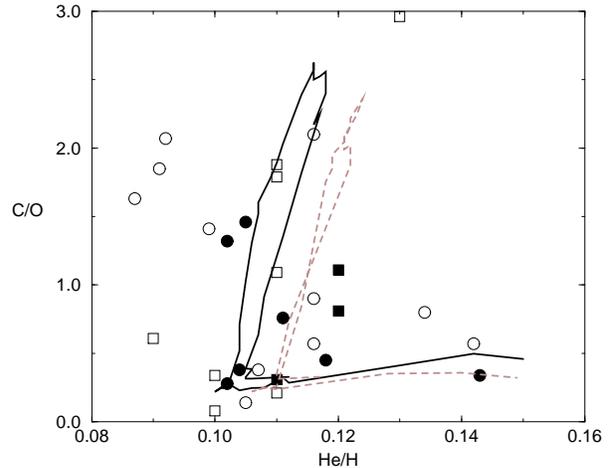}} 
\caption{The symbols have the same meaning as figure
  \ref{fig:heno_comp_all}. The dashed line are the [Fe/H]=0.1 models.}
\label{fig:heco} 
\end{figure}

A full comparison of models to all the data is beyond the scope
of this paper but we show the comparison to N/O vs. He/H and C/O
vs. He/H.  Inspection of Fig.~\ref{fig:heno_comp_all} suggests that
these models fit most of the data reasonably well. We expect the
[Fe/H]=0.0 and 0.1 grids to overlap the majority of the PNe with
N/O$>0.5$ since these are the most massive progenitors that
experience hot--bottom burning. Objects with lower N/O are fit by low
mass models with a large range of [Fe/H].  Figure~\ref{fig:heco} shows
the comparison of models to PNe data on a plot of C/O to He/H. These models
also fit the data as expected.

\subsection{Results}
\begin{figure*}
\resizebox{\hsize}{!}{\includegraphics[angle=-90]{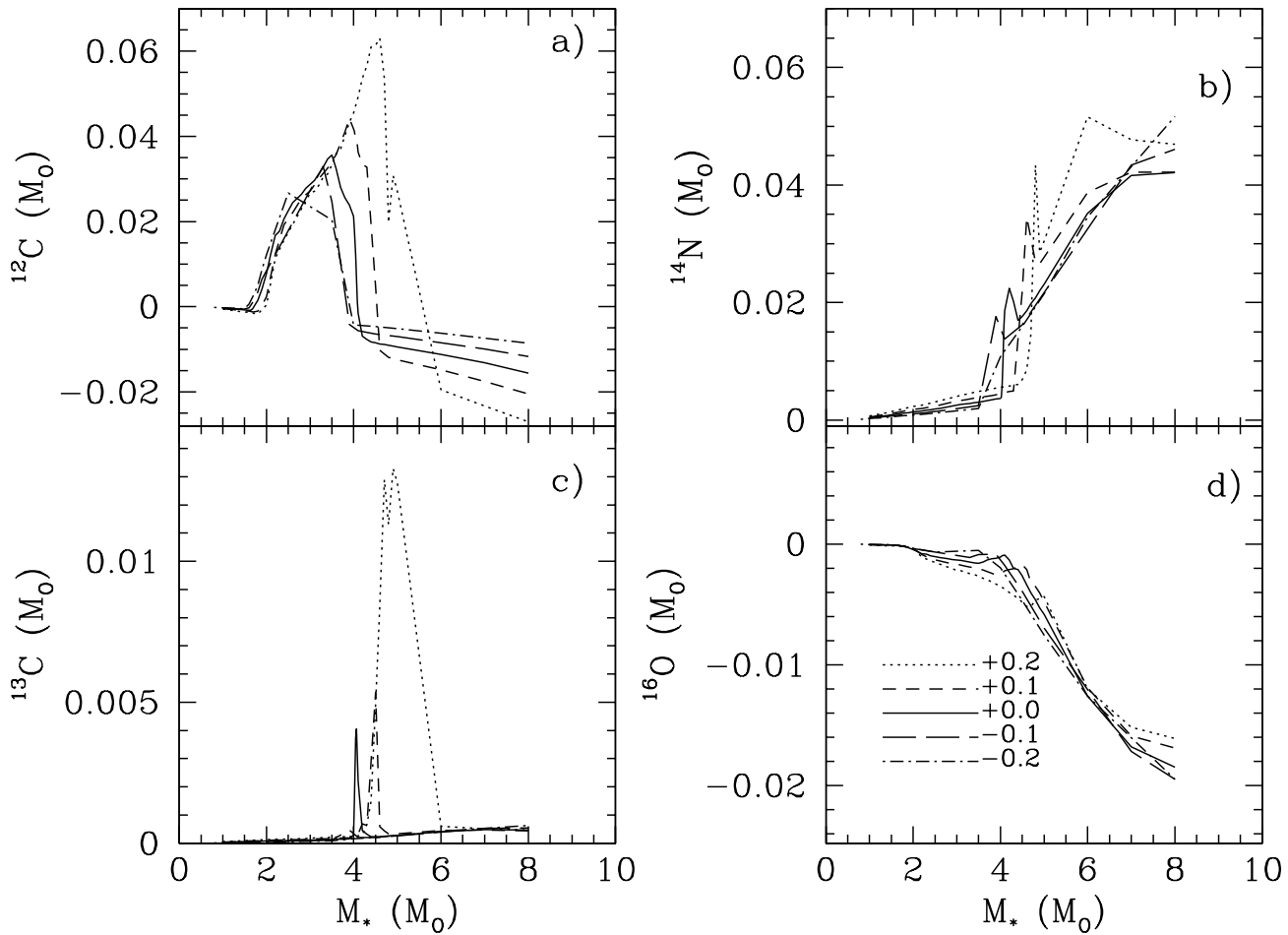}}
\caption{Total yields of $^{12}C$, $^{13}C$, $^{14}N$ and $^{16}O$ 
produced by LIM stars for different metallicities following label 
in panel d, expressed as $log{(Z/Z_{\odot})}$. }
\label{yields}
\end{figure*}

 The resulting stellar yields are given in Table~\ref{yieldsori} for 5
 values of metallicities: $\log{(Z/Z_{\odot})}= -0.2$, -0.1, 0.0, +0.1
 and +0.2, and for masses from 1 to 8 $M_{\odot}$. We show in
 Table~\ref{yieldsori} only the solar metallicity results. The
 complete table with the other metallicity sets is available in
 electronic format. These resulting yields for $^{12}$C, $^{13}$C,
 $^{14}$N and $^{16}$O are represented in Fig.~\ref{yields}.

The yield of $^{12}{\rm C}$ is shown in panel a of
Figure~\ref{yields}.  This is approximately zero for the lowest mass
stars which exhibit no dredge--up events.  As the metallicity is
decreased, stars of lower mass will produce carbon because dredge--up
events occur in lower mass stars at lower metallicities. The sharp
drop--off in the carbon yield occurs when stars get massive enough to
exhibit HBB. This produces a negative yield because the carbon in the
envelope is converted into nitrogen.

 The yield of $^{14}{\rm N}$ (Figure~\ref{yields}, panel b has a
 local maximum at $\sim 3.5--5 M_{\odot}$ and then decreases slightly
 before beginning to increase as a function of stellar
 mass. Significant amounts of nitrogen are produced solely for
 intermediate mass stars because HBB and the 2$^{\rm nd}$ dredge--up
 occur only in stars with $M>3.5--5 M_{\odot}$. The yields at lower
 masses are due to the 1$^{st}$ dredge--up.

HBB produces higher luminosities, larger radii and the mass--loss rate
increases correspondingly in the stars undergoing this process. As a
consequence, stars which experience HBB have shorter TP--AGB lifetimes
compared to those that do not experience HBB.  This shorter TP--AGB
lifetime results in fewer 3$^{\rm rd}$ dredge--up events and less new
carbon mixed into the envelope that can be converted into primary
nitrogen.  The local maximum between 3.5 and 5 $\msun$ is where
significant nitrogen production due to HBB begins. Models below the
mass of this maximum have lifetimes $\sim 10$ times greater than those
above this maximum. The transition occurs at the onset of hot--bottom
burning. The amount of material dredged--up decreases by a large factor
through this transition zone while the amount of hot--bottom burning
increases, producing the local maximum.

The yield of $^{13}{\rm C}$ (Figure~\ref{yields}, panel c has a
maximum around $\sim 4--5 M_{\odot}$, but this element is also produced
in a smaller quantity for stars with masses $M>5--6 M_{\odot}$. This is
mostly a primary component.  The yield of oxygen (Figure~\ref{yields},
panel d is negative in all stars.

We will compare these yields with those of other authors in
the following section, in particular the proportion of secondary
nitrogen produced in each case.

\footnotesize
\begin{flushleft}
\begin{table*}
\begin{tabular}{ccccccccccc}
\hline
\noalign{\smallskip}
Mass &  H  & He & C12 & N14 &O16 & C13 & C12P & N14P &  O16P & C13P  \\
\noalign{\smallskip}
\hline
\noalign{\smallskip}
\multicolumn{11}{c}{$\rm [Fe/H]=0.0$}  \\ 
 1.00 & -0.13E-01 &  0.13E-01 & -0.35E-03 &  0.40E-03 & -0.42E-04 &
0.30E-04 &  0.00E+00 &  0.00E+00 &  0.00E+00 &  0.00E+00 \\ 
 1.10 & -0.14E-01 &  0.14E-01 & -0.42E-03 &  0.50E-03 & -0.51E-04 &  0.36E-04 &  0.00E+00 &  0.00E+00 &  0.00E+00 &  0.00E+00 \\
 1.20 & -0.16E-01 &  0.16E-01 & -0.51E-03 &  0.59E-03 & -0.59E-04 &  0.42E-04 &  0.00E+00 &  0.00E+00 &  0.00E+00 &  0.00E+00 \\
 1.30 & -0.17E-01 &  0.17E-01 & -0.59E-03 &  0.69E-03 & -0.68E-04 &  0.48E-04 &  0.00E+00 &  0.00E+00 &  0.00E+00 &  0.00E+00 \\
  1.40 & -0.17E-01 &  0.17E-01 & -0.67E-03 &  0.78E-03 & -0.76E-04 &  0.53E-04 &  0.00E+00 &  0.00E+00 &  0.00E+00 &  0.00E+00 \\
  1.50 & -0.17E-01 &  0.17E-01 & -0.76E-03 &  0.89E-03 & -0.84E-04 &  0.58E-04 &  0.00E+00 &  0.00E+00 &  0.00E+00 &  0.00E+00 \\
  1.60 & -0.17E-01 &  0.17E-01 & -0.85E-03 &  0.99E-03 & -0.93E-04 &  0.63E-04 &  0.00E+00 &  0.00E+00 &  0.00E+00 &  0.00E+00 \\
  1.62 & -0.17E-01 &  0.17E-01 & -0.87E-03 &  0.10E-02 & -0.94E-04 &  0.64E-04 &  0.00E+00 &  0.00E+00 &  0.00E+00 &  0.00E+00 \\
  1.64 & -0.17E-01 &  0.17E-01 & -0.86E-03 &  0.10E-02 & -0.97E-04 &  0.65E-04 &  0.29E-04 &  0.00E+00 &  0.14E-06 &  0.00E+00 \\
  1.66 & -0.18E-01 &  0.18E-01 & -0.62E-03 &  0.10E-02 & -0.11E-03 &  0.66E-04 &  0.29E-03 & -0.88E-23 &  0.14E-05 &  0.00E+00 \\
  1.68 & -0.18E-01 &  0.18E-01 & -0.56E-03 &  0.11E-02 & -0.11E-03 &  0.67E-04 &  0.36E-03 &  0.00E+00 &  0.17E-05 &  0.00E+00 \\
  1.70 & -0.19E-01 &  0.18E-01 & -0.21E-03 &  0.11E-02 & -0.13E-03 &  0.68E-04 &  0.74E-03 &  0.29E-21 &  0.36E-05 &  0.00E+00 \\
  1.80 & -0.24E-01 &  0.22E-01 &  0.15E-02 &  0.12E-02 & -0.20E-03 &  0.72E-04 &  0.26E-02 &  0.00E+00 &  0.13E-04 &  0.00E+00 \\
  1.90 & -0.32E-01 &  0.26E-01 &  0.41E-02 &  0.13E-02 & -0.31E-03 &  0.76E-04 &  0.53E-02 &  0.00E+00 &  0.27E-04 &  0.00E+00 \\
  2.00 & -0.42E-01 &  0.33E-01 &  0.79E-02 &  0.14E-02 & -0.46E-03 &  0.79E-04 &  0.92E-02 &  0.00E+00 &  0.48E-04 &  0.00E+00 \\
  2.10 & -0.55E-01 &  0.41E-01 &  0.12E-01 &  0.15E-02 & -0.63E-03 &  0.82E-04 &  0.14E-01 &  0.18E-20 &  0.73E-04 &  0.00E+00 \\
  2.20 & -0.69E-01 &  0.50E-01 &  0.17E-01 &  0.15E-02 & -0.80E-03 &  0.84E-04 &  0.18E-01 & -0.81E-18 &  0.98E-04 &  0.00E+00 \\
  2.30 & -0.72E-01 &  0.52E-01 &  0.18E-01 &  0.17E-02 & -0.85E-03 &  0.89E-04 &  0.19E-01 &  0.43E-17 &  0.11E-03 &  0.00E+00 \\
  2.40 & -0.81E-01 &  0.58E-01 &  0.21E-01 &  0.18E-02 & -0.96E-03 &  0.92E-04 &  0.23E-01 &  0.37E-17 &  0.13E-03 &  0.00E+00 \\
  2.50 & -0.86E-01 &  0.61E-01 &  0.22E-01 &  0.19E-02 & -0.10E-02 &  0.95E-04 &  0.25E-01 &  0.90E-18 &  0.14E-03 &  0.00E+00 \\
  2.60 & -0.92E-01 &  0.65E-01 &  0.25E-01 &  0.20E-02 & -0.11E-02 &  0.98E-04 &  0.27E-01 &  0.44E-17 &  0.15E-03 &  0.00E+00 \\
  2.80 & -0.97E-01 &  0.67E-01 &  0.26E-01 &  0.23E-02 & -0.12E-02 &  0.11E-03 &  0.29E-01 & -0.70E-17 &  0.17E-03 &  0.00E+00 \\
  2.90 & -0.10E+00 &  0.69E-01 &  0.28E-01 &  0.24E-02 & -0.13E-02 &  0.11E-03 &  0.30E-01 & -0.14E-19 &  0.18E-03 &  0.00E+00 \\
  3.10 & -0.10E+00 &  0.71E-01 &  0.30E-01 &  0.26E-02 & -0.14E-02 &  0.12E-03 &  0.32E-01 & -0.63E-18 &  0.20E-03 &  0.00E+00 \\
  3.30 & -0.11E+00 &  0.75E-01 &  0.33E-01 &  0.28E-02 & -0.15E-02 &  0.12E-03 &  0.36E-01 &  0.77E-17 &  0.23E-03 &  0.00E+00 \\
  3.40 & -0.12E+00 &  0.79E-01 &  0.35E-01 &  0.29E-02 & -0.15E-02 &  0.13E-03 &  0.38E-01 & -0.12E-17 &  0.25E-03 &  0.00E+00 \\
  3.50 & -0.12E+00 &  0.81E-01 &  0.36E-01 &  0.30E-02 & -0.16E-02 &  0.13E-03 &  0.39E-01 &  0.70E-17 &  0.26E-03 &  0.00E+00 \\
  3.60 & -0.11E+00 &  0.74E-01 &  0.33E-01 &  0.32E-02 & -0.15E-02 &  0.14E-03 &  0.36E-01 &  0.28E-17 &  0.25E-03 &  0.00E+00 \\
  3.70 & -0.97E-01 &  0.64E-01 &  0.28E-01 &  0.33E-02 & -0.13E-02 &  0.15E-03 &  0.31E-01 &  0.41E-17 &  0.24E-03 &  0.00E+00 \\
  3.80 & -0.89E-01 &  0.59E-01 &  0.26E-01 &  0.34E-02 & -0.12E-02 &  0.15E-03 &  0.29E-01 &  0.58E-09 &  0.25E-03 &  0.25E-07 \\
  3.90 & -0.84E-01 &  0.56E-01 &  0.24E-01 &  0.36E-02 & -0.12E-02 &  0.16E-03 &  0.27E-01 &  0.12E-06 &  0.25E-03 &  0.46E-05 \\
  4.00 & -0.76E-01 &  0.51E-01 &  0.21E-01 &  0.37E-02 & -0.11E-02 &  0.29E-03 &  0.25E-01 &  0.41E-05 &  0.24E-03 &  0.10E-03 \\
  4.02 & -0.70E-01 &  0.47E-01 &  0.18E-01 &  0.40E-02 & -0.10E-02 &  0.15E-02 &  0.22E-01 &  0.16E-03 &  0.22E-03 &  0.10E-02 \\
  4.04 & -0.65E-01 &  0.43E-01 &  0.12E-01 &  0.61E-02 & -0.97E-03 &  0.37E-02 &  0.17E-01 &  0.18E-02 &  0.20E-03 &  0.27E-02 \\
  4.06 & -0.59E-01 &  0.39E-01 &  0.67E-02 &  0.10E-01 & -0.94E-03 &  0.41E-02 &  0.13E-01 &  0.47E-02 &  0.19E-03 &  0.27E-02 \\
  4.08 & -0.54E-01 &  0.36E-01 &  0.13E-02 &  0.16E-01 & -0.92E-03 &  0.31E-02 &  0.88E-02 &  0.84E-02 &  0.17E-03 &  0.19E-02 \\
  4.10 & -0.50E-01 &  0.33E-01 & -0.15E-02 &  0.19E-01 & -0.92E-03 &  0.22E-02 &  0.67E-02 &  0.10E-01 &  0.16E-03 &  0.12E-02 \\
  4.20 & -0.37E-01 &  0.25E-01 & -0.69E-02 &  0.22E-01 & -0.13E-02 &  0.45E-03 &  0.30E-02 &  0.11E-01 &  0.12E-03 &  0.15E-03 \\
  4.30 & -0.27E-01 &  0.18E-01 & -0.77E-02 &  0.20E-01 & -0.20E-02 &  0.32E-03 &  0.26E-02 &  0.72E-02 &  0.94E-04 &  0.12E-03 \\
  4.40 & -0.16E-01 &  0.10E-01 & -0.83E-02 &  0.17E-01 & -0.21E-02 &  0.23E-03 &  0.22E-02 &  0.34E-02 &  0.60E-04 &  0.85E-04 \\
  4.50 & -0.30E-01 &  0.25E-01 & -0.85E-02 &  0.18E-01 & -0.26E-02 &  0.23E-03 &  0.22E-02 &  0.32E-02 &  0.60E-04 &  0.92E-04 \\
  4.60 & -0.46E-01 &  0.41E-01 & -0.87E-02 &  0.19E-01 & -0.34E-02 &  0.24E-03 &  0.22E-02 &  0.34E-02 &  0.63E-04 &  0.11E-03 \\
  4.70 & -0.64E-01 &  0.58E-01 & -0.89E-02 &  0.20E-01 & -0.40E-02 &  0.25E-03 &  0.22E-02 &  0.37E-02 &  0.68E-04 &  0.12E-03 \\
  4.80 & -0.81E-01 &  0.74E-01 & -0.91E-02 &  0.21E-01 & -0.46E-02 &  0.26E-03 &  0.23E-02 &  0.40E-02 &  0.74E-04 &  0.13E-03 \\
  4.90 & -0.98E-01 &  0.91E-01 & -0.92E-02 &  0.22E-01 & -0.53E-02 &  0.27E-03 &  0.24E-02 &  0.44E-02 &  0.80E-04 &  0.14E-03 \\
  5.00 & -0.11E+00 &  0.11E+00 & -0.94E-02 &  0.23E-01 & -0.58E-02 &  0.28E-03 &  0.24E-02 &  0.48E-02 &  0.86E-04 &  0.15E-03 \\
  6.00 & -0.28E+00 &  0.26E+00 & -0.11E-01 &  0.35E-01 & -0.13E-01 &  0.41E-03 &  0.27E-02 &  0.87E-02 &  0.16E-03 &  0.21E-03 \\
  7.00 & -0.42E+00 &  0.41E+00 & -0.13E-01 &  0.42E-01 & -0.17E-01 &  0.48E-03 &  0.28E-02 &  0.90E-02 &  0.19E-03 &  0.20E-03 \\
  8.00 & -0.55E+00 &  0.54E+00 & -0.16E-01 &  0.42E-01 & -0.18E-01 &  0.47E-03 &  0.24E-02 &  0.56E-02 &  0.15E-03 &  0.14E-03 \\
\hline
\noalign{\smallskip}
\end{tabular}
\caption{Yields produced in low and intermediate stars for solar metallicity}
\label{yieldsori}
\end{table*}
\end{flushleft}
\normalsize

\section{Comparison with other yields}
\label{sec2}

\subsection{The set of stellar yields used}

Our initial objective is to check the
complete set of metallicity-dependent yields for LIM stars (m $<= 8
\rm M_{\odot}$) obtained as explained in Section 2, using it as the input
in a Galactic chemical evolution model.  First, we would like to
compare the results of the yields presented here with those produced
by the use of other standard yields.

To compare our yields, hereafter BU, we have selected VG and MA ones.
BU yields are calculated for five metallicities, as described, MA are
given for three ($Z =0.004$, 0.008, and 0.02), and the VG yields have
the same metallicity values as MA plus $\rm Z=0.001$ and $\rm
Z=0.04$. For what refers to RV yields, the most widely used set in
this range of mass, since they have been improved by the most recent
cited works we will not use them.  However we compare in this section
our resulting yields with those from RV, too.

\begin{figure*}
\resizebox{\hsize}{!}{\includegraphics[angle=-90]{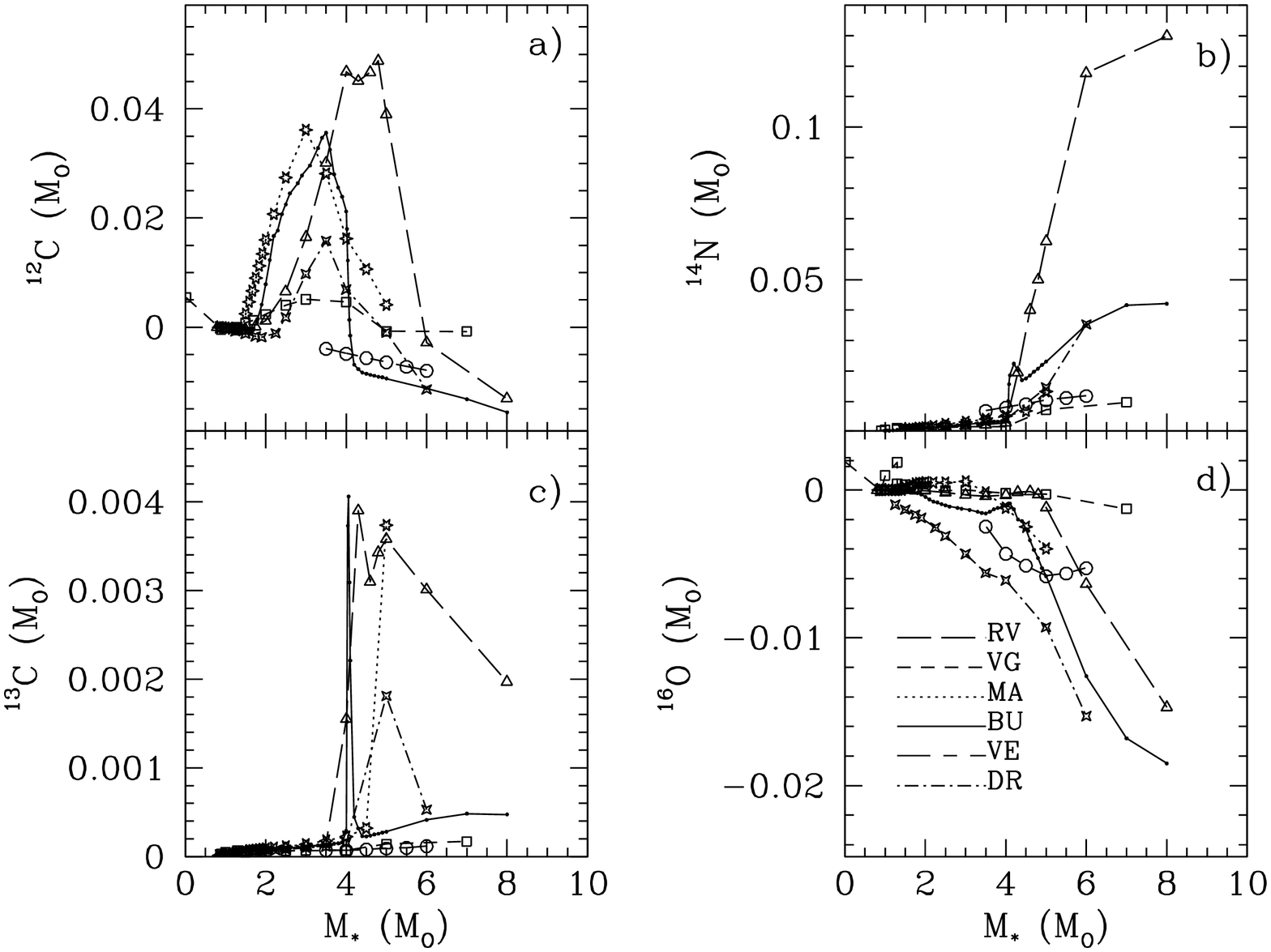}}
\caption{Total yields of $^{12}C$, $^{13}C$, $^{14}N$ and $^{16}O$
produced by LIM stars of solar composition in this work, called BU,
compared with those given by RV, VG, MA, VE and DR marked with
different symbols according panel d. }
\label{compara}
\end{figure*} 

We compare the BU yields corresponding to solar metallicity, with
those of other authors in Fig.~\ref{compara}. There we have
also included the yields recently
computed by \cite{ven02}, VEN (open dots), and by \cite{dray03}, DR
(stars). Although we will not use these two sets, mostly due to their
smaller range of stellar mass and/or Z of the computed models compared
with the ours, we want to check that BU yields are not in
disagreement with those computed with more precise techniques.

We see in panel a, which refers to $^{12}C$, that yields from RV
(model with $\alpha=1.50$ and $\eta=0.33$ of their Table 3d) are the
largest producer of carbon. Yields from MA
(their models for $\alpha=2$) are similar to BU, but, as they are
calculated assuming $M_{up}=5 M_{\odot}$, that is, only for stellar
masses lower than (or equal to) this value, they stop at the mass
for which our yields becomes negative. The final effect is a larger
total yield for MA compared to BU when the integration is done over the
whole LIM stellar range. The values given by VG (their model from
Table 17, solar abundance, $\eta_{AGB}=4$ and $m_{HBB}=0.8 M_{\odot}$)
are lower for stellar masses smaller than 4 $M_{\odot}$, but they
remain positive or almost zero after that, while BU yields
are negative for the most massive LIM stars.  DR also produces
a strong maximum around 3 $M_{\odot}$ in $^{12}C$, although slightly
lower than BU.  Unfortunately, VE only give values for stellar masses
larger than 3.5$M_{\odot}$ which prevents us from checking if the
value of the maximum in BU yields must be decreased, as the DR line
seems to suggest. The behavior of VE after 4$M_{\odot}$ is similar to
the BU and DR models with negative and decreasing values for
increasing stellar masses.

These negative values correspond to the production of nitrogen in the
same range of mass, seen in panel b. It is clear that RV yields also
produce a larger amount of nitrogen than BU.  The differences with
RV are due to the time the models undergo HBB. In our models, which
achieve higher base temperatures than RV, the HBB epoch lasts 
approximately one-tenth of their time.  Therefore, RV models
burn more O into N. On the other hand, MA and VG produce negligible
nitrogen yields for solar metallicity in comparison with RV. BU shows
an intermediate behavior between these two extremes.  The yields from
VEN falls on the VG values, while DR has similar yields to both except
the last point for 6$M_{\odot}$, which is equal to that of BU. In
this case the maximum obtained by BU around 4 $M_{\odot}$ does not
appear in any of the two new yields, VEN or DR.  We think that it is a
problem of sampling, since a smaller step in the computed stellar
masses is necessary to see this effect.

For $^{13}C$, panel c, differences between authors are small: RV, 
BU, MA, and even DR show a large maximum around 4-5
$M_{\odot}$, while VG and VEN results are very low without a maximum. Panel
d shows yields for oxygen, where it is evident that all of them are
negative for $M > 4 M_{\odot}$, as expected, due to the production of
N by the CNO cycle.

We caution about the sampling problem that appears with some yield
sets.  Due to the wide mass step used in the tables, yields have not
always been computed for the mass for which the maximum appears in BU
yields.  Our models will calculate the corresponding yield by
interpolating between these two values but, if a maximum was between
them, it would be missed. This implies that the integrated yield along
the Initial Mass Function (IMF) might be smaller than the true
value. We suggest the use of a smaller step in the range between 3 to
5 $M_{\odot}$ to compute stellar models so as not to lose those
phases of the star that are important and in which the largest amounts
of elements are ejected.

All these differences will have an effect on the final abundances
obtained by the chemical evolution model, as we will see in Section 4.
Besides the differences existing between the published stellar yields,
we have shown that the BU yields are not in disagreement with more
precise techniques used recently.  Due to the shorter computation time
need to obtain BU yields, these could be calculated for a wide range
of stellar masses and Z to see details on the stellar mass dependence,
which are not otherwise seen.  Therefore, until the new techniques are
refined and their corresponding results are available, the BU yields
are accredited to be used in chemical evolution models.

\subsection{The final input: the ejected masses of elements}

We now must add the set of massive stars yields, (m $> 8 \rm M_{\odot}$)
in order to have the whole mass range to include in a chemical
evolution galaxy model. We have chosen the yields of WW and PCB for massive
stars. The choice of WW is made because it is a well--known set used as
a reference by many authors.  We have also chosen PCB yields because
the treatment they give to the mass loss by winds is accurate and
because they use the evolutionary tracks of the Padova Group, that are
widely known.  We do not need to compute models using more yield sets
because they have already been compared in other works. We will 
refer to these other works in Section \ref{discu}.

\footnotesize
\begin{flushleft}
\begin{table*}
\begin{tabular}{lccccccccc}
\hline
\noalign{\smallskip}
M &  Mrem & H & He4 & C & $^{13}$C & N  & $^{13}$C$\rm _{s}$ & N$\rm _{s}$ & O \\
\noalign{\smallskip}
\hline
\noalign{\smallskip}
\multicolumn{10}{c}{$Z=0.02$}\\
  0.80 & 0.53 & 1.81E-01 & 8.30E-02 & 5.80E-04 & 0.00E+00 & 0.00E+00& 2.90E-05  & 4.43E-04 & 2.04E-03\\
  1.00 & 0.56 & 0.30E+00 & 0.14E+00 & 0.90E-03 & 0.00E+00 & 0.00E+00& 0.47E-04  & 0.77E-03 & 0.33E-02 \\
  1.10 & 0.57 & 0.36E+00 & 0.16E+00 & 0.11E-02 & 0.00E+00 & 0.00E+00& 0.56E-04  & 0.93E-03 & 0.40E-02 \\
  1.20 & 0.57 & 0.42E+00 & 0.19E+00 & 0.12E-02 & 0.00E+00 & 0.00E+00& 0.65E-04  & 0.11E-02 & 0.47E-02 \\
  1.30 & 0.59 & 0.49E+00 & 0.21E+00 & 0.14E-02 & 0.00E+00 & 0.00E+00& 0.74E-04  & 0.13E-02 & 0.53E-02 \\
  1.40 & 0.60 & 0.55E+00 & 0.24E+00 & 0.16E-02 & 0.00E+00 & 0.00E+00& 0.82E-04  & 0.14E-02 & 0.60E-02 \\
  1.50 & 0.62 & 0.61E+00 & 0.26E+00 & 0.17E-02 & 0.00E+00 & 0.00E+00& 0.90E-04  & 0.16E-02 & 0.66E-02 \\
  1.60 & 0.63 & 0.67E+00 & 0.29E+00 & 0.19E-02 & 0.00E+00 & 0.00E+00& 0.99E-04  & 0.18E-02 & 0.73E-02 \\
  1.62 & 0.63 & 0.68E+00 & 0.29E+00 & 0.19E-02 & 0.00E+00 & 0.00E+00& 0.10E-03  & 0.18E-02 & 0.75E-02 \\
  1.64 & 0.63 & 0.69E+00 & 0.29E+00 & 0.20E-02 & 0.00E+00 & 0.00E+00& 0.10E-03  & 0.19E-02 & 0.76E-02 \\
  1.66 & 0.63 & 0.71E+00 & 0.30E+00 & 0.23E-02 & 0.00E+00 & 0.00E-00& 0.10E-03  & 0.19E-02 & 0.77E-02 \\
  1.68 & 0.63 & 0.72E+00 & 0.31E+00 & 0.24E-02 & 0.00E+00 & 0.00E+00& 0.11E-03  & 0.19E-02 & 0.78E-02 \\
  1.70 & 0.64 & 0.73E+00 & 0.31E+00 & 0.28E-02 & 0.00E+00 & 0.29E-21& 0.11E-03  & 0.20E-02 & 0.79E-02 \\
  1.80 & 0.65 & 0.79E+00 & 0.34E+00 & 0.47E-02 & 0.00E+00 & 0.00E+00& 0.11E-03  & 0.21E-02 & 0.86E-02 \\
  1.90 & 0.65 & 0.85E+00 & 0.37E+00 & 0.76E-02 & 0.00E+00 & 0.00E+00& 0.12E-03  & 0.23E-02 & 0.92E-02 \\
  2.00 & 0.66 & 0.90E+00 & 0.40E+00 & 0.12E-01 & 0.00E+00 & 0.00E+00& 0.13E-03  & 0.25E-02 & 0.97E-02 \\
  2.10 & 0.67 & 0.96E+00 & 0.44E+00 & 0.16E-01 & 0.00E+00 & 0.18E-20& 0.14E-03  & 0.26E-02 & 0.10E-01 \\
  2.20 & 0.68 & 0.10E+01 & 0.47E+00 & 0.21E-01 & 0.00E+00 & 0.00E-18& 0.14E-03  & 0.28E-02 & 0.11E-01 \\
  2.30 & 0.68 & 0.11E+01 & 0.50E+00 & 0.22E-01 & 0.00E+00 & 0.43E-17& 0.15E-03  & 0.30E-02 & 0.11E-01 \\
  2.40 & 0.69 & 0.11E+01 & 0.53E+00 & 0.25E-01 & 0.00E+00 & 0.37E-17& 0.15E-03  & 0.32E-02 & 0.12E-01 \\
  2.50 & 0.70 & 0.12E+01 & 0.56E+00 & 0.28E-01 & 0.00E+00 & 0.90E-18& 0.16E-03  & 0.34E-02 & 0.13E-01 \\
  2.60 & 0.71 & 0.12E+01 & 0.58E+00 & 0.30E-01 & 0.00E+00 & 0.44E-17& 0.17E-03  & 0.35E-02 & 0.13E-01 \\
  2.80 & 0.73 & 0.14E+01 & 0.64E+00 & 0.32E-01 & 0.00E+00 & 0.00E-17& 0.18E-03  & 0.40E-02 & 0.15E-01 \\
  2.90 & 0.73 & 0.14E+01 & 0.66E+00 & 0.34E-01 & 0.00E+00 & 0.00E-19& 0.19E-03  & 0.42E-02 & 0.15E-01 \\
  3.10 & 0.75 & 0.16E+01 & 0.72E+00 & 0.36E-01 & 0.00E+00 & 0.00E-18& 0.20E-03  & 0.46E-02 & 0.16E-01 \\
  3.30 & 0.77 & 0.17E+01 & 0.77E+00 & 0.40E-01 & 0.00E+00 & 0.77E-17& 0.22E-03  & 0.49E-02 & 0.18E-01 \\
  3.40 & 0.78 & 0.17E+01 & 0.80E+00 & 0.42E-01 & 0.00E+00 & 0.00E-17& 0.22E-03  & 0.51E-02 & 0.18E-01 \\
  3.50 & 0.79 & 0.18E+01 & 0.83E+00 & 0.43E-01 & 0.00E+00 & 0.70E-17& 0.23E-03  & 0.53E-02 & 0.19E-01 \\
  3.60 & 0.81 & 0.19E+01 & 0.84E+00 & 0.40E-01 & 0.00E+00 & 0.28E-17& 0.24E-03  & 0.55E-02 & 0.20E-01 \\
  3.70 & 0.82 & 0.19E+01 & 0.86E+00 & 0.36E-01 & 0.00E+00 & 0.41E-17& 0.25E-03  & 0.57E-02 & 0.21E-01 \\
  3.80 & 0.84 & 0.20E+01 & 0.87E+00 & 0.34E-01 & 0.25E-07 & 0.58E-09& 0.26E-03  & 0.59E-02 & 0.21E-01 \\
  3.90 & 0.87 & 0.21E+01 & 0.89E+00 & 0.32E-01 & 0.46E-05 & 0.12E-06& 0.27E-03  & 0.60E-02 & 0.22E-01 \\
  4.00 & 0.89 & 0.21E+01 & 0.91E+00 & 0.30E-01 & 0.10E-03 & 0.41E-05& 0.30E-03  & 0.62E-02 & 0.23E-01 \\
  4.02 & 0.88 & 0.21E+01 & 0.91E+00 & 0.27E-01 & 0.10E-02 & 0.16E-03& 0.59E-03  & 0.64E-02 & 0.23E-01 \\
  4.04 & 0.87 & 0.22E+01 & 0.92E+00 & 0.21E-01 & 0.27E-02 & 0.18E-02& 0.12E-02  & 0.70E-02 & 0.23E-01 \\
  4.06 & 0.86 & 0.22E+01 & 0.92E+00 & 0.16E-01 & 0.27E-02 & 0.47E-02& 0.14E-02  & 0.81E-02 & 0.23E-01 \\
  4.08 & 0.86 & 0.22E+01 & 0.92E+00 & 0.10E-01 & 0.19E-02 & 0.84E-02& 0.13E-02  & 0.99E-02 & 0.24E-01 \\
  4.10 & 0.85 & 0.22E+01 & 0.93E+00 & 0.75E-02 & 0.12E-02 & 0.10E-01& 0.11E-02  & 0.11E-01 & 0.24E-01 \\
  4.20 & 0.86 & 0.23E+01 & 0.94E+00 & 0.25E-02 & 0.15E-03 & 0.11E-01& 0.42E-03  & 0.14E-01 & 0.24E-01 \\
  4.30 & 0.87 & 0.24E+01 & 0.96E+00 & 0.19E-02 & 0.12E-03 & 0.72E-02& 0.33E-03  & 0.16E-01 & 0.24E-01 \\
  4.40 & 0.89 & 0.25E+01 & 0.98E+00 & 0.15E-02 & 0.85E-04 & 0.34E-02& 0.27E-03  & 0.16E-01 & 0.25E-01 \\
  4.50 & 0.90 & 0.25E+01 & 0.10E+01 & 0.16E-02 & 0.92E-04 & 0.32E-02& 0.27E-03  & 0.17E-01 & 0.25E-01 \\
  4.60 & 0.91 & 0.26E+01 & 0.11E+01 & 0.16E-02 & 0.11E-03 & 0.34E-02& 0.26E-03  & 0.18E-01 & 0.25E-01 \\
  4.70 & 0.91 & 0.26E+01 & 0.11E+01 & 0.17E-02 & 0.12E-03 & 0.37E-02& 0.27E-03  & 0.19E-01 & 0.25E-01 \\
  4.80 & 0.92 & 0.27E+01 & 0.11E+01 & 0.18E-02 & 0.13E-03 & 0.40E-02& 0.27E-03  & 0.20E-01 & 0.25E-01 \\
  4.90 & 0.93 & 0.27E+01 & 0.12E+01 & 0.19E-02 & 0.14E-03 & 0.44E-02& 0.28E-03  & 0.21E-01 & 0.25E-01 \\
  5.00 & 0.93 & 0.28E+01 & 0.12E+01 & 0.20E-02 & 0.15E-03 & 0.48E-02& 0.28E-03  & 0.22E-01 & 0.25E-01 \\
  6.00 & 0.99 & 0.33E+01 & 0.16E+01 & 0.28E-02 & 0.21E-03 & 0.87E-02& 0.39E-03  & 0.31E-01 & 0.25E-01 \\
  7.00 & 1.04 & 0.38E+01 & 0.20E+01 & 0.35E-02 & 0.20E-03 & 0.90E-02& 0.50E-03  & 0.38E-01 & 0.29E-01 \\
  8.00 & 1.08 & 0.43E+01 & 0.24E+01 & 0.38E-02 & 0.14E-03 & 0.56E-02& 0.60E-03  & 0.42E-01 & 0.34E-01 \\
\hline
\noalign{\smallskip}
\end{tabular}
\caption{Ejected mass by the LIM stars during their evolution. We only
present results for LIM stars, solar metallicity and elements until
oxygen.  The complete table with the whole range in mass and metallicity 
 is only available in electronic format.}
\label{ejections}
\end{table*}
\end{flushleft}
\normalsize 

Thus, for massive stars we will use WW and PCB while we have
used three set of yields for LIM stars: those presented here, VG and
MA. We have combined them, obtaining 3 models: 1) BU-WW; 2) VG-WW;
3) MA-PCB, which we call BU, VG and MA.

Some authors produce yields while others give their
results as ejections, which are not equivalent. The stellar yield of
an element is the amount that has been newly created and ejected
during the evolution of the star, while the ejection computes not
only this new mass of the element but the original one, which
corresponds to the initial metallicity of the star, too.  The yield
can be negative, if the star transforms more of the element
than it creates, but the ejection is always positive.  The formula to
transform one into the other \citep{tin80} is:

\begin{equation}
E_{i}=Y_{i}+(M_{ini}-M_{rem}) X_{i}
\end{equation} 

where $E_{i}$ is the ejected mass of the element i, $Y_{i}$ is the
value of the yield for the same element, $M_{ini}$ is the initial mass
of the star, $M_{rem}$ is the remnant mass and $X_{i}$ is the original
stellar abundance of the i-element.

The different ways of presenting the data should be taken into account
when the input values of the model for the whole mass range are
constructed. WW and PCB give their results as ejected masses for each
element while RV, MA and BU directly produce the stellar yields, so we
have to transform all them into the same type of quantity before using
them as code inputs. We follow the formalism of PCB for the matrix
elements $Q_{i,j}$, as we describe in the next section, so we prefer
to work with ejections, in order to directly apply their equations. In
the present work, we have included both the stellar yields in
Table~\ref{yieldsori}, already described, and the corresponding
ejections in Table~\ref{ejections}.

Thus, we show in this last table the complete ejected mass set for
some metallicities. The mass
ejected for each element is given by each stellar mass from
$m_{low}=0.8 M_{\odot}$ to $m_{sep}=8 M_{\odot}$ for solar
metallicity.  Column (1) is the stellar mass, column (2) is the
remnant mass, columns (3) to (10) show the ejected mass of $^{12}C$,
$^{13}C$, $^{14}N$, $^{16}O$, and $^{12}C_{P}$, $^{13}C_{P}$,
$^{14}N_{P}$, $^{16}O_{P}$.  These values result from the 
computed yields shown in Table 1. The complete table, given only in
electronic format, provides results for metallicities
$log(Z/\zsun)=-0.2$,-0.1,+0.0,+0.1 and +0.2, by including the massive
star yields, for masses up to $m_{up}=100 M_{\odot}$. This implies
that columns (3) to (10) correspond to CNO elements, as explained, but
columns from (11) to (16) show the
ejected masses for other elements produced only by massive stars: Ne,
Mg, Si, S, Ca and Fe. For massive stars these quantities are
obtained from the production factors by WW. A similar table has been
computed for the two other models, VG and MA.

In all cases we use metallicity--dependent yields. However, the range
of metallicity for which the yields are given is different for each
set. It has been necessary to transform them into homogeneous and
consistent sets for which we have adapted the values of the
metallicity of massive and LIM stars to a common set in each
model. When the metallicities are not the same for the massive star set
as for the LIM star yields, as occurs with models BU and VG with respect
to WW, we need to interpolate in Z to obtain a complete table for each
Z. In that case, we have preferred to interpolate the massive
star yields to compute them at the metallicities given for LIM stars,
because they have a more continuous variation in Z for the whole mass
range than LIM stars.  Thus, for instance, for BU we have the
yields for $ Z=0.0159$, and we interpolate those from WW between
$Z=0.1Z_{\odot}$ and $Z=Z_{\odot}$ at this same abundance value.  We
use the same technique for the two other sub--solar metallicities
(-0.2,-0.5), by interpolating between tables corresponding to
$Z=Z_{\odot}$ and Z=0.1Z$_{\odot}$. For the over--solar sets, and
because that WW do not give yields for metal--rich stars,
we have extrapolated their solar yields to use with the yields of
$Z=0.025$ and $Z=0.0317$ from BU.  For the VG model we
also interpolate in the massive stars tables to obtain the complete
tables at the abundances given by VG, from $Z=0.001$ to $Z=0.04$.

When we use MA with PCB yields the problem is smaller because both
sets are calculated for a similar set of abundances. However, MA do
not give results for the lowest abundances in the Padova group, 
$Z=0.0004$. We have analyzed if the trend in Z is clear and
continuous, and as this does not occur for $^{14}N$, as we will see
later, we have preferred to use the smallest metallicity yields from
MA ($Z=0.004$) along with tables for $Z=0.0004$ from PCB.

\section{Galactic Chemical Evolution Models}\label{results} 

\subsection{Summary description of the multiphase chemical evolution model} 

The model used in this work is the Multiphase Chemical Evolution Model
described in \citet{fer92,fer94}, in the version presented
in \citet{mol04}.  The Galaxy
is considered as a two--zone system: halo and disk, sliced into
cylindrical concentric regions. It calculates the time evolution of
five different populations or matter phases in the Milky
Way: diffuse gas, molecular clouds, low and intermediate stars,
massive stars and stellar remnants.

The corresponding yields for type Ia and Ib supernova explosions, 
included in calculations following \cite{tor89} and \cite{fer93}, are
taken from \cite{iwa99} and \cite{bra86}.  The assumed initial mass
function is from \cite{fer90}, very similar to the one given by
\cite{kro01}.

The chemical abundances of 15 elements are computed through the
Q--matrix technique. The Q--matrix is based on the \cite{tal73}
formalism and well described in previous publications of this code.
Each element of the matrix, $Q_{i,j}$ is the mass fraction of the star
initially in the form of species $j$ that has been transformed to
species $i$ and ejected. The original formalism changes for the
metallicity--dependent yields. Thus, we have taken the equations given
by PCB for all elements except for D and $^{3}He$, for which the
relations given by \cite{gal95} are used. To compute the Q--matrix we
use the tables with the ejected masses of elements computed as
described in the previous section.

\subsection{Calibration: The solar neighborhood}

\begin{figure}
\resizebox{\hsize}{!}{\includegraphics[angle=0]{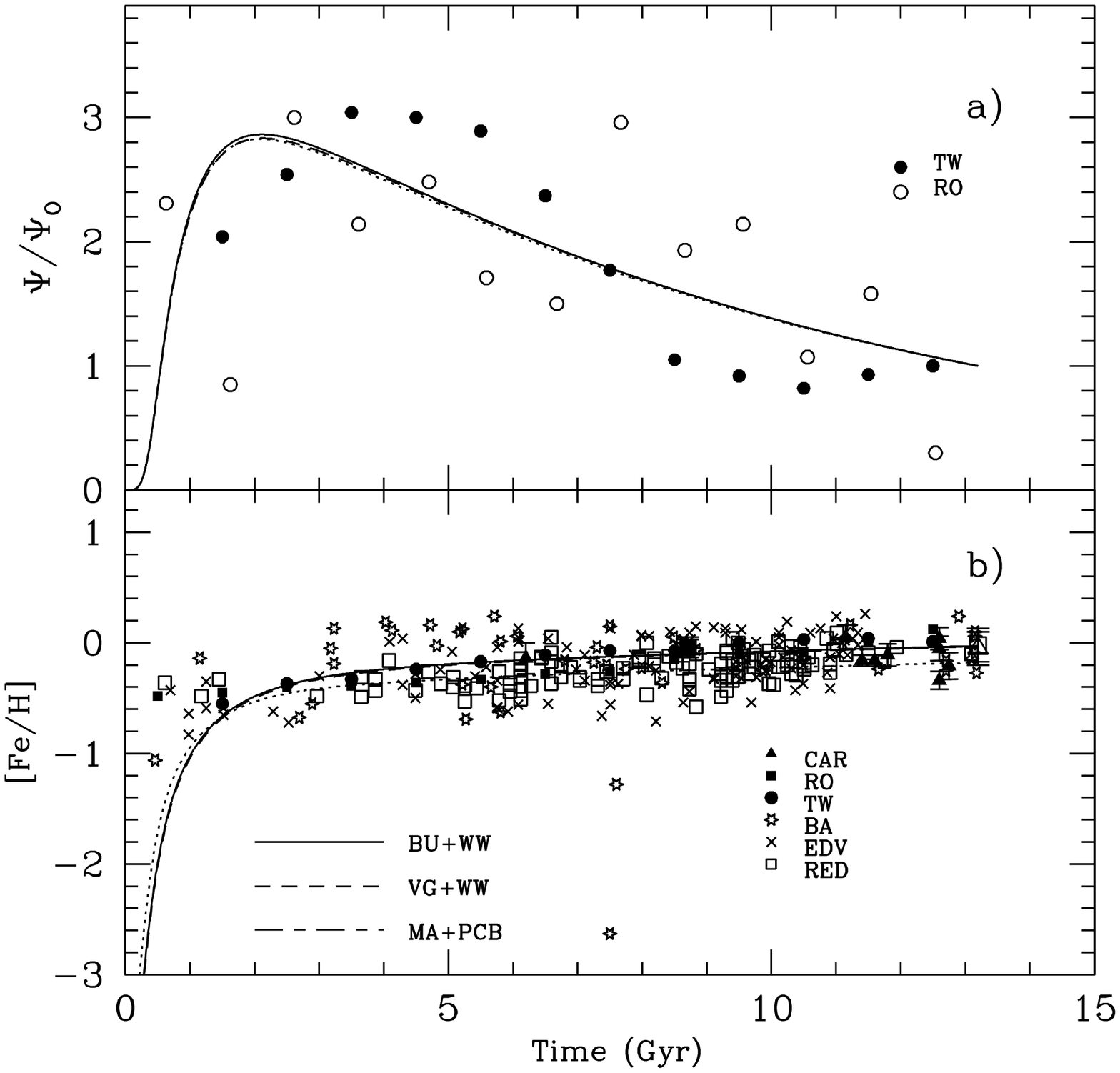}}
\caption{Time evolution in the Solar Vicinity: a) the star formation
history and b) the age--metallicity relation. Data are from
\cite{twa80} (TWA), \cite{ba88} (BA), \cite{edv93} (EDV), \cite{car98}
(CAR), \citet{rocha00,rocha00b} (RO) and \cite{red03}.}
\label{sv}
\end{figure}

We now use the three described sets of yields in the GCE Model. The
model was already used in the Solar Vicinity (SV), assumed as the
region located at a galactocentric distance of 8 kpc, and in the
Galactic disk, \citep{fer92,fer94}, so we do not need to again compare
atomic and molecular gas, or star formation rate radial distributions
with data \citep[ but see][ for a revised comparison of these observational
constraints with the model]{mol04}.

However, as we are using a new version of the code where
metallicity--dependent yields have been implemented, we have to check
that model results are still correct. On the other hand, as our
objective is to compare the resulting carbon abundances given by
different stellar yields, only other relations where these element
yields have no influence must be analyzed.  Thus, in this section we
only use model results independent of the selected yield set.  So, we
choose the SFR history, the age--metallicity relation and the G--dwarf
metallicity distribution, all of them for the Solar Vicinity, a
well--known region where the number of observational constraints is
high.

In Fig.~\ref{sv} we have: a) the star formation history and b) the
age--metallicity relation for the SV.  Our 3 models, BU, VG, and MA,
are represented by solid, short--dashed, and dotted lines
respectively. The star formation history recently obtained by
\citet{rocha00b} shows a behavior more similar to the one obtained
from hydrodynamical simulations of galaxy formation \citep{sai03},
with strong variations with the time compared to the smoother line
from the \citet{twa80} data. Nevertheless, the average star formation
history is similar, with a maximum around 3 Gyr and another, more
recent, around 10 Gyr.

The age--metallicity relation, which mostly depends on SN--I iron
ejection, is almost the same for all models.  Some differences are
apparent however, because a quantity of iron is also ejected by
massive stars. The PCB yields produce more iron t iron than those from WW,
and therefore model MA shows a higher metallicity at early times than
models BU and VG.  In order to better reproduce the present abundance
data, the rate of SN--Ia was decreased in model MA, compared to
WW. Due to this the iron will appear later in WW models than in the
model MA. This will have an effect on our final results.

\begin{figure}
\resizebox{\hsize}{!}{\includegraphics[angle=0]{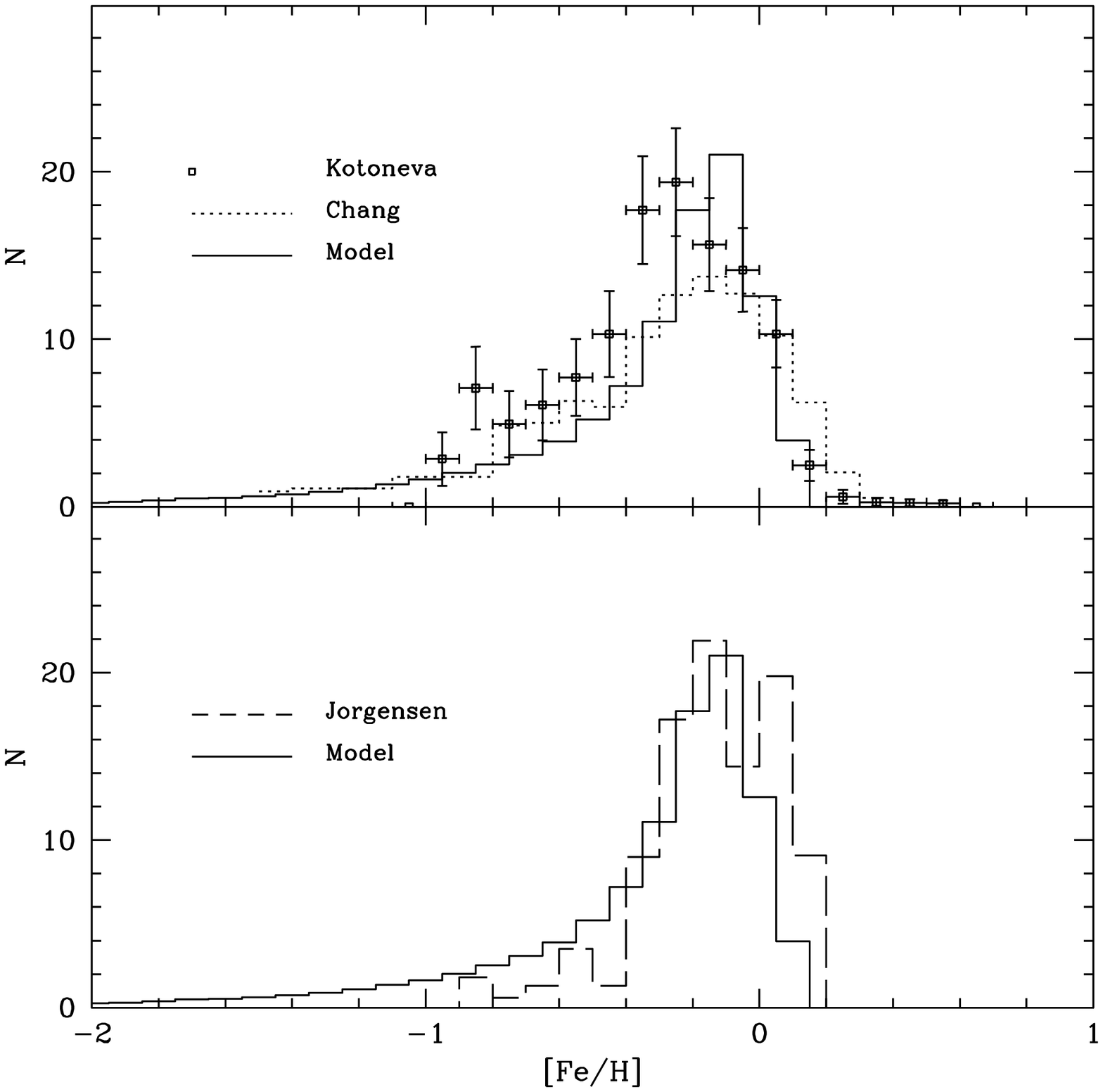}}
\caption{G--dwarf distribution for our  model compared with data from
\citet{chan00,kot02} in a) and with data from \cite{jor00} in b).}
\label{gd}
\end{figure}

The G--dwarf metallicity distribution is shown in Fig.~\ref{gd} for our
 model BU and compared with data from \cite{jor00}. We see that the
 model produces iron with a maximum around -0.10 dex, very similar to
 the observed average which is $\sim 0.16$ dex.  This distribution is
 much more peaked (or less wide) than those found in others works
 \citep{pag89,chan00,rocha00} and similar to the one from
 \citet{kot02}. These last authors confirm that the solar vicinity
 formed over a long time scale, of the order of 7 Gyr, similar to 
ours. Taking into account the dispersion of measurements, estimated
 by the variations shown by these different sets of data, the modeled
 distribution may be considered acceptable. Since iron comes mostly
 from SN--Ia, the resulting distribution for the model MA is similar
 which for the sake of clarity we do not show.

Once a suitable calibration of the model is obtained, we can analyze
the differences of carbon abundances resulting from the cited sets of
yields.  

\subsection{The time evolution of CNO abundances} 

Data used for the comparison in the next figures are taken 
following Table~\ref{authors}.  We start with the time
evolution in the Solar Neighborhood.

\begin{figure*}

\resizebox{\hsize}{!}{\includegraphics[angle=0]{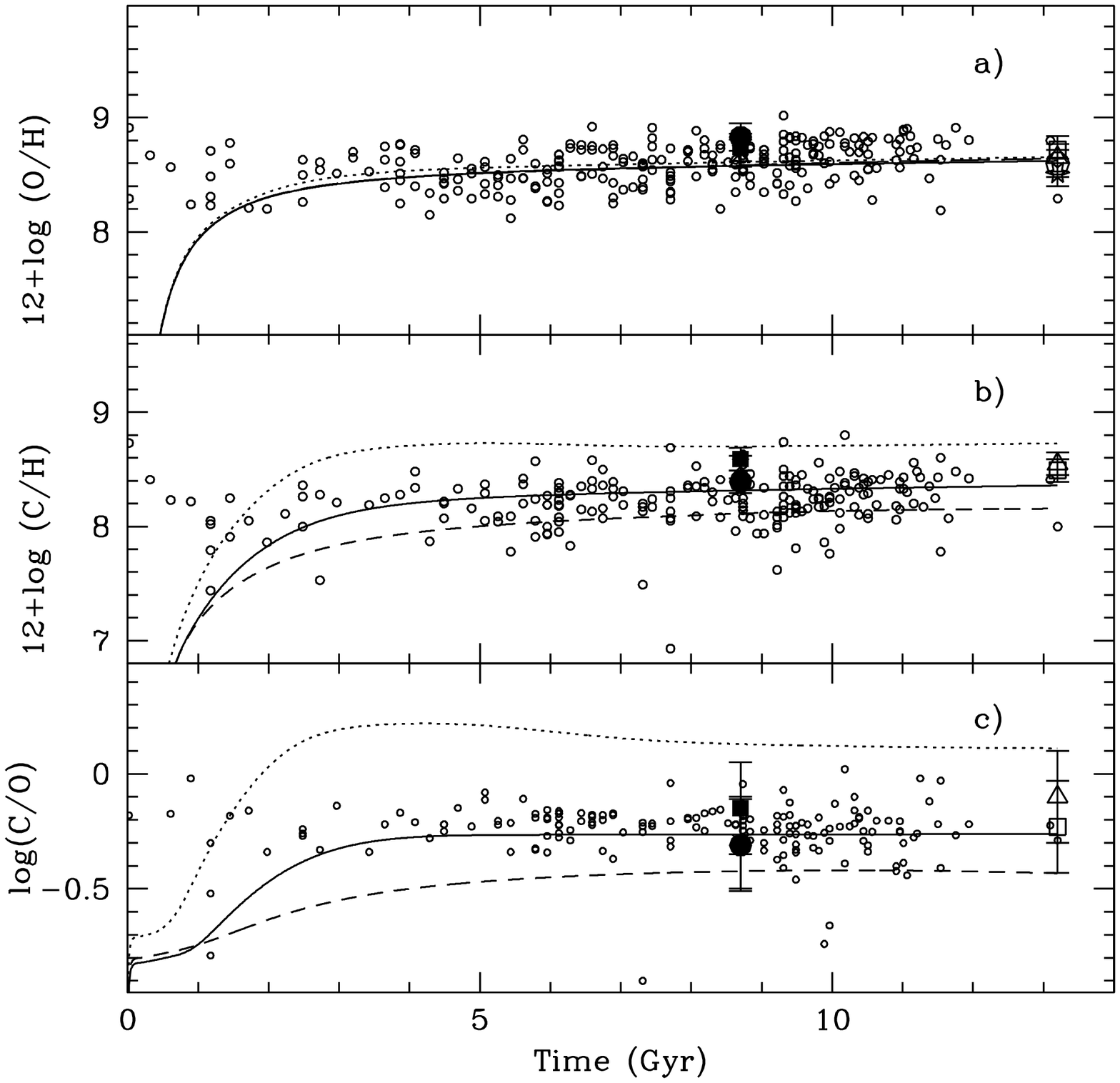}}

\caption{Time evolution of abundances in the Solar Vicinity: a)
oxygen, b) carbon, as $12+log(X/H)$ and c) the ratio C/O. Solar
abundances are the filled symbols from \cite{gre98} -- circles--,
\cite{hol01} --squares-- and \cite{all01,all01b,all02}, --crosses--,
by assuming an age of 4.5 Gyr for the sun.  Empty symbols at 13.2 Gyr
are the interstellar medium abundances given by \cite{mey97,mey98},
--circles--, \cite{pei99} --squares--, \cite{sofia01}, --triangles--,
and \cite{moos02}, --stars. Small open dots are stellar abundances
obtained from Table~\ref{authors}, being those located 
around the Solar Vicinity ($7 < R < 9$ kpc). Line meanings are given in
panel b of Fig.~\ref{sv}.}
\label{abunt}
\end{figure*}

Fig.~\ref{abunt} shows the evolution of oxygen --panel a-- carbon
--panel b-- and the relative abundance log(C/O) --panel c--.
Models are represented by the same symbols as in
Fig.~\ref{sv}. There are almost no differences between models for
oxygen, all of them being in agreement with the data. Since oxygen
proceeds mostly from massive stars, it is obvious than models BU and
VG produce equal abundances. Model MA gives a slightly higher
abundance (O/H) that corresponds to a larger yield in PCB than in WW. 

%with the solar abundance values --large filled symbols-- taken from
%\cite{gre98} -- circles--, \cite{hol01} --squares-- and
%\cite{all01,all01b,all02}, --crosses--, by assuming an age of 4.5 Gyr
%for the sun. For the interstellar medium abundances at 13.2 Gyr, large
%empty symbols, we use the abundances given by \cite{mey97,mey98},
%--circles--, \cite{pei99} --squares--, \cite{sofia01}, --triangles,
%and \cite{moos02}, --stars. The small circles are data of stars
%located at galactocentric distances between 7 and 9 kpc.

\footnotesize
\begin{flushleft}
\begin{table*}
\begin{tabular}{cccccc}
\hline
\noalign{\smallskip}
[Fe/H] & [C/H]  & [O/H]    & R     & Age     & Reference \\
\noalign{\smallskip}
\hline
\noalign{\smallskip}
  X      &  X  &   X  & --- & --- & \cite{ake03}\\
  X      & --- &   X  & --- & --- & \cite{bar88}\\
  X      & --- &   X  & --- & --- & \cite{bar89}\\
  X      & --- &   X  & --- & --- & \cite{boe99}\\
  X      & X   & ---  & --- & --- & \cite{car87}\\
  X      & X   &   X  & --- & --- & \cite{carr00}\\
  X      & --- &   X  & --- & --- & \cite{cav97}\\
  X      & --- &   X  & --- &  X  & \cite{chen00}\\
  X      & X   &   X  & --- & --- & \cite{cle81}\\
  X      & X   &   X  & --- & --- & \cite{daf01}\\
  X      & X   &   X  & --- & --- & \cite{dep02}\\
  X      & --- &   X  & X   & X   & \cite{edv93}\\
 ---     & X   &  --- & X   & X   & \cite{gus99}\\
  X      & X   &  --- & --- & --- & \cite{fri90}\\
  X      & X   & X    & --- & --- & \cite{gra00}\\
 ---     & X   & X    & X   & --- & \cite{gum98}\\
 X       & --- & X    & --- & --- & \cite{isr98}\\
         &     &      &     &     & \cite{isr01}\\
 X       &  X  & ---  & --- & --- & \cite{lai85}\\
 X       &  X  & ---  & --- & --- & \cite{mel01}\\
         &     &      &     &     & \cite{mel02}\\
 X       & --- &  X   & --- & --- & \cite{mis00}\\
 X       & --- &  X   & --- & --- & \cite{niss02}\\
         &     &      &     &     & \cite{niss02b}\\
 X       & X   & ---  & --- & X   & \cite{red03}\\
 X       & X   & X    & X   & --- & \cite{rol00}\\
         &     &      &     &     & \cite{sma97}\\
         &     &      &     &     & \cite{sma01}\\
 X       & X   & ---  & --- & --- & \cite{shi02}\\
 X       & --- & X    & --- & --- & \cite{smi01}\\
 X       & X   & X    & --- & --- & \cite{tomk84}\\
         &     &      &     &     & \cite{tomk86}\\
         &     &      &     &     & \cite{tomk95}\\
 X       & X   & X    & --- & --- & \cite{wes00}\\
\hline
\noalign{\smallskip}
\end{tabular}
\caption{References for CNO stellar abundances used for the comparison
with model results.}
\label{authors}
\end{table*}
\end{flushleft}
\normalsize

Differences among models in the carbon abundances are larger than the
error bars for the Solar and ISM data: Model BU is located in the lower
part of the error bars, while the MA model is above the data. VG shows the
lowest values. WW yields do not produce as much carbon as PCB, because
the stellar winds, which mostly ejected this element, are not
considered in WW. Due to this, MA is the highest in panel
b. Models BU and VG are coincident in the first Gyr, when carbon
proceeds from the same massive stars.  After a time $\sim$ 1.5
Gyr, a difference appears between these two models, indicating a
greater carbon production by the BU yields. The final consequence is
a better fit of the observations by the model BU.

Similar information can also be extracted from the relative abundances
{\sl vs} time in Fig.~\ref{abunt}. The fact that carbon is fit well by
the BU--WW yields, without needing massive star yields incorporating
mass loss and its dependence on metallicity, probably indicates that
the mass loss assumed in massive star yields different to WW is too
high.

Several zones are clearly distinguished in panel a as in panel b:
first, an abrupt increase which corresponds to the massive star
contribution. Afterwards, an almost constant value indicates the region
where the bulk of stars between 4 to 8 $M_{\odot}$,  which do
not eject carbon, are dying. Then, intermediate mass stars ($M\sim 3-5M_{\odot}$)
begin to eject carbon, producing an amount sufficient to reach the
present value. This can be seen clearly in this figure, because the point where
the models begin to separate is where the LIM stars carbon
contribution appears.  \citet{car00} points out that the time evolution
shown by this graph seems to indicate an increase of the carbon
abundance at recent times and that metallicity--dependent yields are
necessary to explain this finding. However, we see that the data may be reproduced
by an almost constant evolution of C/O after the increase shown
between $\sim $1 and 3 Gyr. An increase for the youngest objects ($\rm
t > 10$ Gyr) is not apparent. The number of
observations has increased and we think that probably the difference
in conclusions is due to these more recent stellar data.

\subsection{Radial gradients of abundances} 

\begin{figure*}
\resizebox{\hsize}{!}{\includegraphics[angle=0]{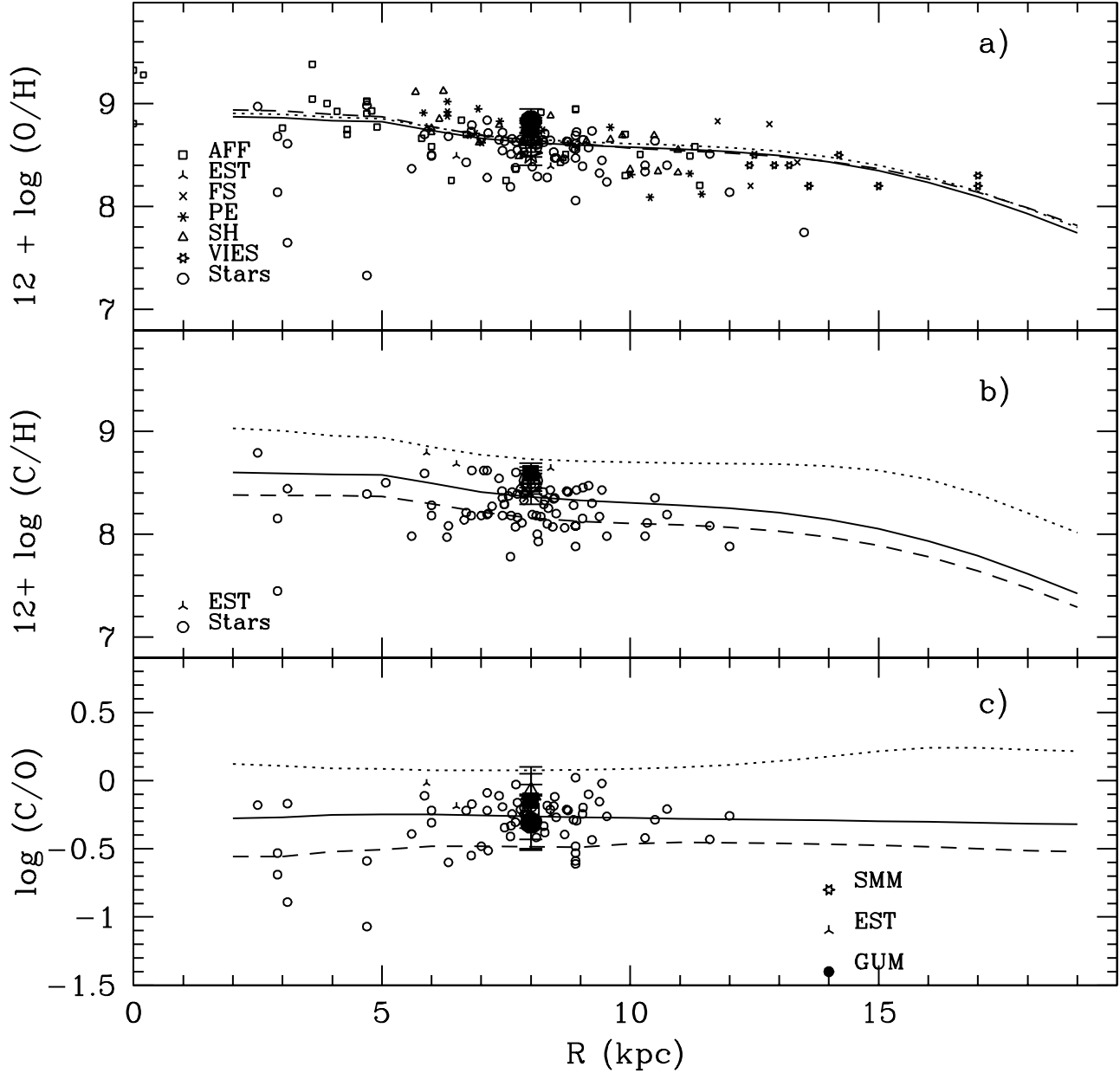}}
\caption{Radial distributions of elemental abundances, as 12 + log
(X/H), in the Galaxy disk for: a) oxygen, b) carbon and the relative
abundance log(C/O). Stellar data --filled dots-- are taken from
authors given in Table~\ref{authors}.  H{\sc ii} regions abundances
are also included, taken from \cite{pei79} --PE--, \cite{sha83}
--SHA--, \cite{fs91} --FS--, \cite{vies96} --VIES--,
\cite{aff97}--AFF-, and \cite{este99,este99b,este99c} --EST-- as
marked in the figure.}
\label{abunr}
\end{figure*}

Up to now we have only analyzed SV data, but we want to extend the
results to the entire disk. We plot the radial
gradient of oxygen and carbon as can be seen in Fig.~\ref{abunr}. The 
H{\sc ii} regions data authors are given in the graph while star data are
from Table \ref{authors}.

These radial distributions show a correct shape in the three
panels. This occurs because the shape depends mostly on the infall/SFR
ratio along the galactocentric radius, and it is not
yield--dependent. The value of the gradient at the center and at the
outer regions is currently a matter of discussion \citep{vies96}.

We see in panel a that average abundances of oxygen are even in the
inner disk, where the distribution flattens. The three model radial
distributions lie between the error bars defined by the dispersion of
data.  In panel b, we show the same kind of graph for the carbon
abundance, as $\rm 12+log(C/H)$.  As we would expect, the MA model
presents higher values than two others and does not reproduce the
observations. The two other models fit the data reasonably well.

In panel c of the same figure the relative abundance C/O radial
distribution is plotted.  In agreement with previous comments, the MA
model presents a carbon excess, so it is located outside the data
region. We note that the observed C/O radial
distribution, after including data from \cite{sma01} and \cite{red03},
does not show a negative slope, as \cite{car00} suggested.

\subsection{The relative abundances of carbon} 

The relative abundance of an element gives important clues about how
the evolution is taking place. In this kind of figure, model
parameters like infall rate and star formation efficiency have a
smaller influence than in others.

\begin{figure}
\resizebox{\hsize}{!}{\includegraphics[angle=-90]{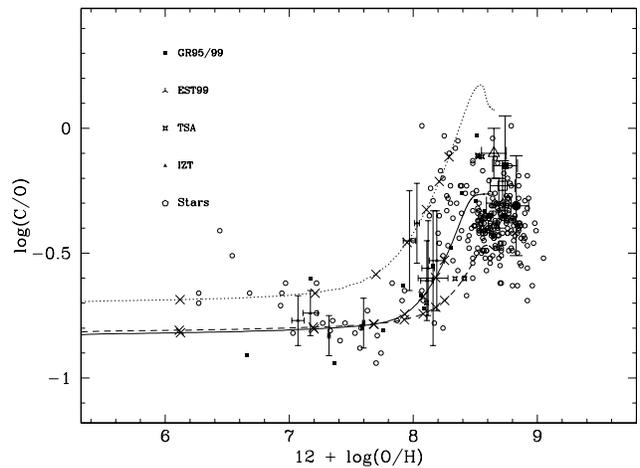}}
\caption{ The relative abundances of log (C/O) {\sl vs} the oxygen
abundance as $12+log(O/H)$.  Stellar data --open circles-- are from
Table~\ref{authors}. Other data are from \cite{este99} --EST99--,
\cite{tsa03}--TSA--, for galactic H{\sc ii} regions, from
\cite{gar95,gar99} --GR95/99--, for other galaxies H{\sc ii} regions
and from \cite{izo99} --IZ-- for Blue compact dwarf galaxies.  The
meaning of symbols is given in the figure. } 
\label{co}
\end{figure}

Fig.~\ref{co} plots the relation between C/O for the SV region
compared with stellar and H{\sc ii} region data. As we are showing the
SV results it would be adequate to compare them with data of the same
region but unfortunately some data could belong to other radial
regions. The ones from \cite{sma01} correspond to the inner disk, and
other sets \citep{edv93,cle81,gum98} may include stars at different
galactocentric distances than the assumed one for the SV. This may
partially be the reason for the large data dispersion usually seen in
this kind of figure.  We try to select data for stars or H{\sc ii}
regions located in the Solar Neighborhood, that is 7 kpc $\rm \le R
\le$ 9 kpc, in order to decrease the dispersion, but the
galactocentric distance is not always given in the sets of data.

We are showing the results representing the evolution of the {\sl
disk}, not that of the halo. In our scenario, both regions are
followed separately, and, as the halo regions have very low star
formation efficiencies, their evolution is not similar to that in the
disk regions. In order to compare with observations, we must choose
only disk data. Usually, all low metallicity objects are considered
halo objects, but this is not always true. To properly select data,
some kinematic information for the stars is necessary. We will use all
possible stellar data although some may come from halo stars.

Following the same argument, observations from other galaxies are not
adequate too, since their evolution may follow other tracks, 
because the star formation or infall rate histories are not
necessarily equal to the SV region \citep{mol96,mol99}. As differences 
in this kind of figure are mainly due to yields, not to
the model parameters, and due to the paucity of carbon data in the
solar neighborhood, we will use data from other parts of the disk, or
even from the halo.

The initial value of log (C/O) between -0.7 and -0.8 dex is caused by
the evolution of massive stars in the first Myr. The flat left part of
the graph must be interpreted having in mind that it takes place in a
very short time: the time needed for stars with masses between 5 and 8
$M_{\odot}$ to evolve.  Stars in this range do not eject carbon, so
the carbon abundance level remains at the level due to the massive
stars ejection, whose contribution appears before.  When stars with
masses close to 4 $M_{\odot}$ begin to die, the carbon increases
rapidly and finally reaches a plateau, when the smallest stars evolve
without ejecting CNO cycle elements.

This interpretation is valid for all models, and to clearly show our
argument crosses have been marked on the model lines to indicate the
times corresponding to 0.25, 0.50, 0.75, 1.0, 1.5 and 2 Gyr since time
zero of the galaxy evolution. The model predicts a continuous star
formation history with a maximum at $\sim 1.8$ Gyr. Most of stars
create in this period and more massive than 4 $M_{\odot}$) (whose
lifetime is $\tau= 0.165$ Gyr) no longer exist at $t=2$ Gyr.

The BU model reaches final values closer to the observations. The MA model
gives a final log(C/O) slightly higher, due to the large production of
carbon in PCB for massive stars. This production is caused by the high
mass loss assumed for the stellar winds, and the comparison between
model and data may imply that the assumed mass loss is too strong.  

Now, we will analyze the abundances {\sl vs} the iron abundances.  Once
again we cannot separate disk and halo objects, so we consider that
disk objects are those with [Fe/H] greater than -1.5. This method is not 
as useful for discriminating between halo and
disk stars as the use of kinematic information, and some halo
stars will be included while some other disk objects of low
metal content will be missed.

\begin{figure*}
\resizebox{\hsize}{!}{\includegraphics[angle=0]{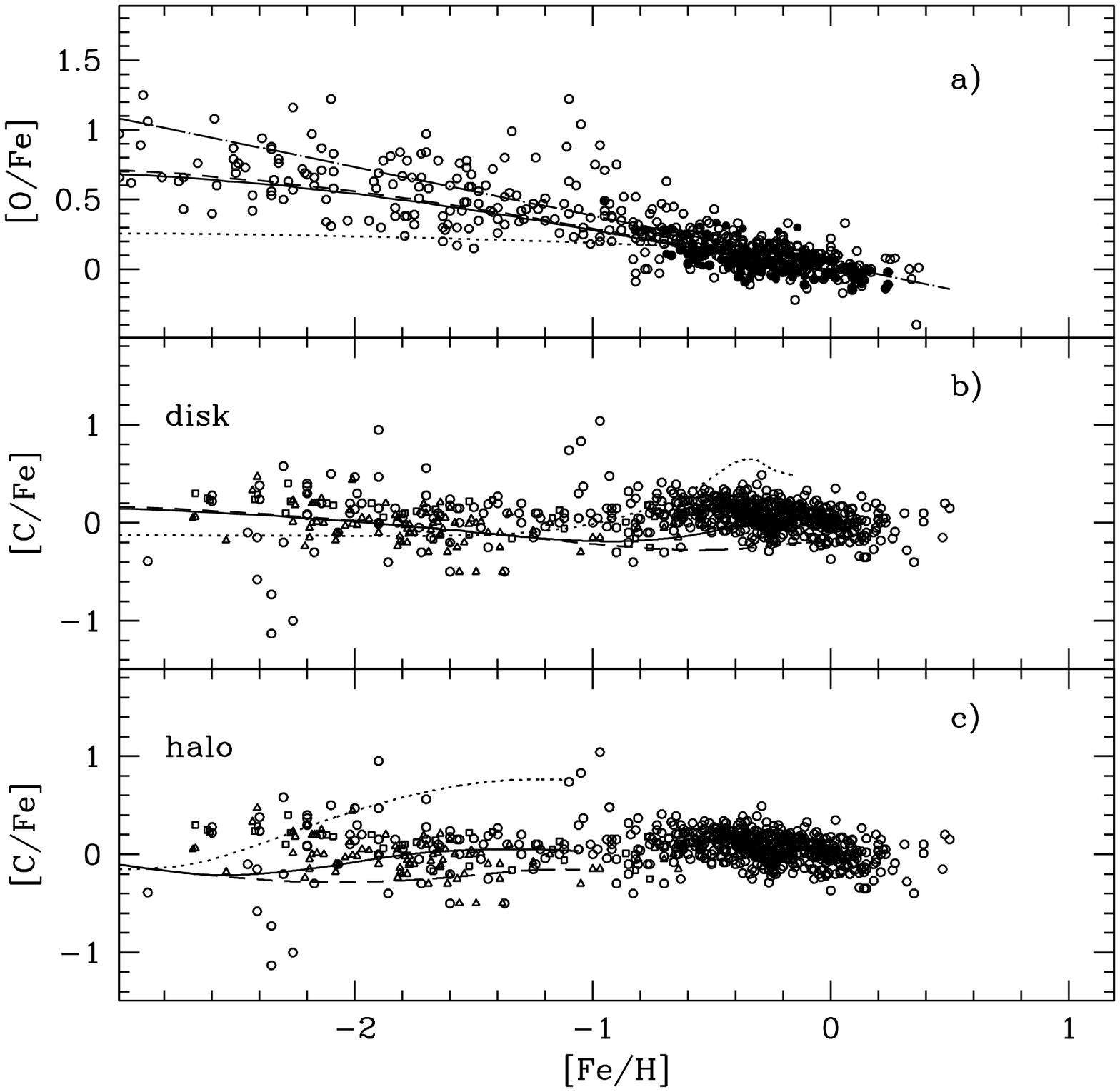}}
\caption{The relative abundances of C and O as [X/Fe] {\sl vs} the
iron abundance [Fe/H]. Panel a shows the [O/Fe] evolution for the
disk. The dot--long-- dashed line represents the trend of \cite{boe99}
data. Panels b and c represent the evolution of [C/Fe] for the disk
and the halo, respectively.  Models have the same line coding as
previous figures. Data are taken from authors from
Table~\ref{authors}.  Open dots are the available data,
while solid dots are stars that have known galactocentric distance in the
range $7 \le R \le 9$ kpc. Triangles are data from \cite{car87}. }
\label{cofe}
\end{figure*} 

The following graph depends not only on the carbon ejection but also
on the iron production. The iron appears mostly as consequence of the
SN--I explosions, which eject a large quantity of this element in each
event. However, massive stars also produce Fe and differences among
the various stellar yields for these stars have an effect on
the results. PCB produce more Fe than WW, and, correspondingly,
[X/Fe] will be smaller for PCB yields when the abundances are low,
while Fe proceeds from massive stars, independently of the yields used
for LIM stars. When the SN--I start to explode, the iron appears in the
ISM and [X/Fe] begins to decrease.  In order to obtain the same iron
at the present time, the SN--Ia rate for the model MA is smaller than
the one from models using WW.  But this only will be seen at later
times. For the early times, the differences may be as high as +0.3 dex
in the abundance of [Fe/H].

The usual correlation between [O/Fe] and [Fe/H] is shown in
Fig.~\ref{cofe}a. Two different trends are usually obtained depending
on the technique used to estimate the stellar abundances. The
dot--long--dashed line represents the trend given by \cite{boe99},
steeper than the second one which shows a flatter shape with
metallicity for $\rm [Fe/H] < -1$. Actually, if the complete set of
data is plotted, the two trends are indistinguishable, as we can see
in panel a, although with a large dispersion. Models BU and VG are in
agreement with these observations while model MA shows a flatter
behaviour at low metallicity.

We represent the relative abundance [C/Fe] {\sl vs} [Fe/H] for the
disk, panel b, and for the halo, panel c, in the same figure.  We have
used the classical data for [C/Fe], but we remind the reader of our
previous argument about the halo and disk as separate entities with
different time evolutions. Some data increase with decreasing
metallicity, and others seem located at a lower level, around 
$\rm [C/Fe] \sim ~0$.

We see in panel b that models BU and VG start with a similar evolution
as it corresponds to the same set of massive stars yields used. For
the disk, both first decrease when metallicity increases. Then, when
the contribution of stars with $M \sim 4 M_{\odot}$ begins to appear,
the BU model increases and has a bump, like the observed one. In fact,
it is difficult to see the model line over the data. Finally, it
decreases when the lowest mass stars begin to die without ejecting any
elements. This model reproduces well the trend described by open dots
at $ [Fe/H] \ge -1$ \citep[and also the data from][]{car87}.  Model
VG, however, continues to decrease after this metallicity, in
disagreement with disk data.  Model MA produces more iron than the two
other models early on, so the absolute [C/Fe] level is lower, than the
two other models, around -0.1 dex, until $\rm [Fe/H] =-1$. However it
produces too much carbon for $[Fe/H]>-1$.  Due to the large dispersion
of data, it is difficult to determine which model (BU or VG) better
fits the observations, but we think that the BU model is more adequate
to fit the disk observations.

The data at low metallicity seem to show two trends shown by
open bullets, and by open triangles, which are the
observations by \cite{car87}. We suggest that the sets represent the
evolution of the halo and the disk, respectively, which are not equivalent. 
We show the evolution of the halo in panel c for the three models.
In this figure there are no lines above $[Fe/H]> -1$ as expected for
the halo. It is clear that MA does not reproduce the low metallicity
observations.  Models BU and VG give results falling in the
regions of these data.  In particular, in this low metallicity region,
the halo of model BU fits well the open dot data while the disk
evolution is closer to the triangles

Therefore, we think that these two different trends correspond to the
different star formation histories occurring in the halo and in the
disk. The BU models are the only ones able to predict the
two observed trends in these data sets.

\section{Discussion}\label{discu} 

We have presented new LIM yields and we have used Galaxy chemical
evolution models to compare the results with data and with
results obtained with other sets of yields.  When carbon abundance data
are analyzed in a graph of log(C/O) {\sl vs} (O/H) an almost flat
slope appears and then they show an increase until the solar
value. This is usually interpreted as the carbon being ejected mostly
by massive stars, thus producing a constant proportion of
C/O. However, the final abrupt increase of the carbon abundances with
increasing oxygen abundance is unexplained.

\cite{car00} invokes the metallicity dependent yields, due to its
effect on mass loss by stellar winds, as essential for solving this
problem of obtaining an increase in the ratio C/O in recent
times. The effect of the winds, which change with Z, included in the
Maeder yields would be able to produce the recent
increase. But the relation (C/O) {\sl vs} (O/H) is not reproduced by
any model (see \cite{car00} Fig.5), in particular the variable slope of
log(C/O) with increasing O/H, because the model produces C/O higher
than data for the low metallicity region.

\cite{hen00} also assume that carbon must proceed mostly from massive
stars and use Maeder yields, but however, had to adjust the carbon
yields by multiplying them by a factor of 3 in order to achieve a good
fit to solar data. This seems to be in contradiction to the hypothesis
that stellar winds proceeding from massive stars explain this
increase.  CHIA03, using yields from \cite{thie96}, also multiply the
carbon yields by a factor of 3 so as to reproduce the present time C
abundance. Thus, it seems impossible to reproduce the observed trend
only with the contribution of massive stars.  Therefore, the best
yields to reproduce carbon abundances seem to be those from WW. When
those authors used WW yields with those from VG, they obtained a
result for the C/O very similar to the one from our VG model.

On the other hand, when the most recent data from carbon abundance are
used, the increasing trend in the abundance of carbon at recent times,
an important argument in \cite{car00} conclusions, seems to disappear,
showing only a large dispersion around a mean C/O mostly flat for
$t > 3$ Gyr.  So, the increase of carbon with respect to the first base
level did not occur in the last 2--3 Gyr but much before.

C/O begins to increase when the oxygen abundance is $12+log(O/H) \sim
8.2$, but this does not necessarily mean that it has occured recently.
In our models this value is reached at around 1.2 Gyr or more than 10
Gyr ago.  Then, when $12+log(C/H) \sim 7.4$, the stars of $\sim 4
M_{\odot}$ begin to contribute to the carbon, which is clear in
Fig.\ref{abunt}b, when models BU and VG, with the same massive star
yields, start to separate. The LIM stars which eject carbon have
masses between 3 and 5 M$_{\odot}$, and have lifetimes around
0.15--0.4 Gyr, short enough in the evolution of the Galaxy. Therefore,
the increase of carbon seen in the relation C/O {\sl vs} O/H is due to
the low mass star ejections. It starts to occur between 1.2 and 3.5
Gyr, or, equivalently, 10 to 12 Gyr ago. CHIA03 claim, on the basis of
their model results, that a large amount of carbon must originate in
LIM stars. Our results are in agreement with this statement. However,
the increase of C/O with metallicity is better reproduced when the new
set of yields BU is used in combination with that of WW for massive
stars.  The yields presented here produce results in excellent
agreement with data.

Thus, we agree with the conclusion from CHIA03 that the evolution of
LIM stars with massive stars without stellar winds may account for the
carbon evolution: the contribution of LIM stars to the final abundance
of carbon may be sufficient to reproduce the observations.  The C
ejected by stars with masses around 4 $M_{\odot}$ produces a strong
and steep increase in the high abundance region starting around
12+log(O/H)$\sim 8.2$. The resulting curve reproduces the observed
trend, in a way not produced by other models.

[C/Fe], instead of C/O, may be analyzed with chemical evolution
models. In \cite{lia01}, who also compare the effect of using several
combinations of yields, all figures refer to Fe and not to O.  Our
model MA has results similar to those shown by \cite{lia01} (model MA
+ PCB) and also by PCB, although the LIM star yields are slightly
different \citep[they use][ instead of MA]{mar96} in this last
model. All of them produce a value of [C/Fe] almost constant up to
$\rm [Fe/H]=-0.5$ that then increases reaching values higher than
observations at the solar metallicity end.

Similarly, CHIA03 also show their results as [C/Fe] {\sl vs}
[Fe/H]. These authors, like \cite{lia01} find a very flat behavior in
this graph with a slight increase at the high abundance end, that the
authors interpret as caused by a significant amount of carbon being
ejected by LIM stars.  Both works show that Model VG is the best one
at reproducing the data corresponding to carbon. Nevertheless, \cite{lia01}
explained that it is possible to fit the carbon data only with the LIM
stars and with stars of $\rm M< 40 M_{\odot}$, without needing carbon
ejected by Wolf--Rayet stars.  Our models BU and VG reproduce a very
similar trend although BU better fits the
data for the high metallicity region, as we see in Fig.~\ref{co}
and Fig.~\ref{cofe}, due to its larger integrated carbon yield.

Since we do not need to invoke a massive star stellar wind effect to
reproduce the observed trend in the C/O or C/Fe data, we think that
the mass loss for massive stars assumed in works such as \cite{mae92}
or PCB is too strong. We are in agreement with \cite{lia01} supporting
a mass loss lower than assumed in previous works, in the same line as
\cite{crow01} who points out that the recent mass--loss rates for
Galactic W--R stars indicate a downward revision of a factor of 2--4
compared with the previous calibrations.

We also support the claim of those authors that the
$^{12}C(\alpha,\gamma)^{16}O$ rate used by WW seems to be
adequate. \cite{hash95} use the rate from \cite{cau85}(CFHZ85), a
factor $\sim$ 2.4 higher than the most recent rate from \cite{cau88}
(CF88), thus producing less carbon. \cite{hash95} gives a discussion
about this rate, indicating that the rate of CF88 is clearly too low,
and produces too much carbon. He claims that, actually, the rate of
CFHZ85, a factor 2.4 larger than that of CF88, is more consistent with
observations, although they stated that an intermediate value, such as
1.7 times the value of CF88, the one used by \cite{woo95}, is probably
better for this reaction rate.

Thus, although VG yields produce reasonable results, BU yields, with
higher absolute values for carbon yields, and following the same
trends as VG, better predicts carbon abundances.

\section{Conclusions} 

\begin{enumerate} 

\item We can reproduce the carbon abundances and the trend of C/O over 
O/H, in particular its increase at almost solar oxygen abundances,
by the effect of LIM stars, without invoking the metallicity
dependence (proceeding the stellar mass loss) of carbon
yields. This conclusion is in agreement with
\cite{lia01,chia03}, and \cite{chia03b}.  The difference to these
last works is that BU yields shown here, with a larger production of
carbon than VG, seem to better fit the data of this element.

\item The previous conclusion implies that the mass loss by stellar
winds in massive stars probably needs to be smoother than usually
assumed. In this sense, we support to \cite{lia01} \citep[but see]{mey02} 
who also claim that the mass loss must be revised downward.

\item The reaction rate for $^{12}C(\alpha,\gamma)^{16}O$ taken as 1.7
times the value of CF88 by WW produces results consistent with the
observations.

\item The data [C/Fe] for $\rm [Fe/H]> -1.0$ is well reproduced by the
evolution of the disk of the BU model.  For low metallicities, the
observations may be divided into two trends.  One of them is well
reproduced by our disk model results, while the second one is well
fitted by the same BU model with the halo results.  These conclusions
are in agreement with recent results from \cite{chia03}, who also show
that the halo and the disk have different evolutions.

 \end{enumerate} 

\begin{acknowledgements} 
This work has been partially supported by
Ministerio de Ciencia y Tecnolog\'{\i}a project AYA--2000--0973. We
thank the referee Dr. F. D'Antona useful comments and suggestions that
improved this work.  
\end{acknowledgements}

\end{document}